\documentclass[
  journal=pasa,
  manuscript=article-type,
  year=2026,
]{cup-journal}

\usepackage{amsmath}
\usepackage{amssymb}
\usepackage[nopatch]{microtype}
\usepackage{booktabs}
\usepackage{dblfloatfix} 
\usepackage{aas-macros}

\usepackage{float}
\usepackage{hyperref}
\usepackage[font=small,skip=0pt]{caption}
\usepackage{subcaption}
\usepackage[utf8]{inputenc}
\usepackage{graphicx} 
\usepackage{xcolor}

\usepackage{pifont} 
\newcommand{\cmark}{\ding{51}}      

\usepackage{fancyhdr}
\title{SM-Net: Learning a Continuous Spectral Manifold from Multiple Stellar Libraries}

\author{Omar Anwar}
\affiliation{International Centre for Radio Astronomy Research (ICRAR)\\ The University of Western Australia (UWA), 35 Stirling Hwy, Crawley WA 6009, Australia}
\email[Omar Anwar]{omar.anwar@uwa.edu.au\ }

\author{Aaron S. G. Robotham}
\affiliation{International Centre for Radio Astronomy Research (ICRAR)\\ The University of Western Australia (UWA), 35 Stirling Hwy, Crawley WA 6009, Australia}

\author{Luca Cortese}
\affiliation{International Centre for Radio Astronomy Research (ICRAR)\\ The University of Western Australia (UWA), 35 Stirling Hwy, Crawley WA 6009, Australia}

\author{Kevin Vinsen}
\affiliation{International Centre for Radio Astronomy Research (ICRAR)\\ The University of Western Australia (UWA), 35 Stirling Hwy, Crawley WA 6009, Australia}

\begin{document}


\begin{abstract}
We present \textsc{SM-Net}, a Machine Learning model that learns a continuous spectral manifold from multiple high-resolution stellar libraries. SM-Net generates stellar spectra directly from fundamental stellar parameters such as effective temperature ($T_{\mathrm{eff}}$), surface gravity ($\log g$), and metallicity ($\log Z$), trained on a combined grid derived from \textsc{PHOENIX--Husser}, \textsc{C3K--Conroy}, \textsc{OB--PoWR} and \textsc{TMAP--Werner} libraries. By combining their respective parameter spaces to achieve broader coverage, we construct an extended composite dataset that spans a significantly larger, more continuous region of stellar parameter space than any single library. The unified grid spans $T_{\mathrm{eff}} = 2{,}000$–$190{,}000$~$\mathrm{K}$, $\log g = -1$ to $9$, and $\log Z = -4$ to $1$, with spectra covering the wavelength range 3{,}000–100{,}000~\AA.

Within the combined parameter space, SM-Net provides smooth interpolation across heterogeneous library boundaries. Outside the sampled domain, it can generate numerically smooth exploratory predictions, although these extrapolated spectra are not directly validated against reference models. Zero or masked flux values are treated as unknowns rather than physical zeros, allowing the network to infer missing regions using correlations learned from neighbouring grid points. Across 3,538 training and 11,530 test spectra, SM-Net achieves a Mean Squared Error (MSE) of
$1.47\times10^{-5}$ on the training set, and $2.34\times10^{-5}$ on the test set respectively, in the transformed log1p-scaled flux representation. These results confirm that the learned manifold reproduces the spectra with high accuracy while remaining robust to incomplete or masked data. Inference throughput exceeds $14,000$ spectra per second on a single GPU. This throughput enables rapid exploration of the stellar parameter space and efficient construction of synthetic stellar population models.

To promote accessibility and community use, we make the SM-Net model publicly available, along with an interactive dashboard that enables real-time spectral generation and visualisation through a web interface. Users can explore the influence of $T_{\mathrm{eff}}$, $\log g$, and $\log Z$ interactively, without a terminal interface, thereby demonstrating a deployed research model ready for exploration. SM-Net is a fast, robust, and flexible data-driven complement to traditional stellar population synthesis libraries, providing a pathway for next-generation stellar population modelling.
\end{abstract}

\section{Introduction}

Stellar spectral libraries are fundamental tools in astrophysics, underpinning analyses from stellar atmosphere studies to stellar 
population synthesis and galactic archaeology. Libraries such as PHOENIX \citep{husser2013new}, C3K \citep{conroy2018metal}, OB-PoWR \citep{hainich2019powr} and 
TMAP \citep{werner2003asp} provides high-resolution synthetic spectra across wide ranges of effective temperature ($T_\mathrm{eff}$), surface 
gravity ($\log g$), and metallicity ($\log Z$). These spectra are generated by stellar-atmosphere models that solve the radiative transfer equation together with the atmospheric structure equations using iterative numerical schemes \citep{hubeny2010atmospheres}. Differences in the physical assumptions, numerical methods, and adopted line lists used in these 
radiative-transfer computations mean that each library spans a particular region of parameter space, with only partial overlaps between them. Consequently, no single library 
provides complete or continuous coverage, and their grids often contain irregular sampling, discontinuities, and masked wavelength regions 
arising from incomplete line lists or numerical instabilities. Transitions between libraries can also introduce non-physical changes in 
spectral features. Any approach aiming to generate continuous spectra must therefore handle missing information, avoid propagating artefacts 
from unreliable grid points, and remain stable near boundaries where library assumptions differ.

Stellar spectral libraries are used in two closely related but conceptually distinct contexts, both of which impose stringent requirements on the quality and continuity of the underlying spectral grids.
In one, they are employed for \emph{direct stellar modelling}, in which observed stellar spectra are compared with library templates to infer atmospheric parameters such as $T_{\mathrm{eff}}$, $\log g$, and metallicity.
In the other case, libraries serve as intermediate building blocks for constructing \emph{simple stellar population} (SSP) models, in which stellar spectra are combined with isochrones, initial mass functions, and evolutionary prescriptions to generate integrated spectra for composite stellar systems.
Widely used frameworks in this latter category include \textsc{BC03} \citep{bruzual2003stellar}, \textsc{Flexible Stellar Population Synthesis (FSPS)} \citep{conroy2010propagation}, \textsc{Binary Population and Spectral Synthesis (BPASS)} \citep{stanway2018re}, and more recently \textsc{ProGeny} \citep{robotham2025progeny}.
While these approaches target different scientific questions, they share a common dependency on the internal consistency, completeness, and smoothness of the underlying stellar spectral grids.

As a consequence of incomplete coverage, heterogeneous assumptions, and irregular sampling in existing libraries, traditional workflows rely on coarse-gridding or interpolation in sparsely populated regions. Interpolation becomes increasingly uncertain when library assumptions change, sampling is irregular, or flux is masked. This limits applications requiring dense 
coverage in $(T_\mathrm{eff}, \log g, \log Z)$, such as stellar population synthesis, Bayesian spectral fitting, or large parameter 
sweeps in stellar evolution modelling.

On the fitting side, tools such as \textsc{pPXF} \citep{cappellari2004parametric, cappellari2017improving} treat 
libraries as fixed inputs and determine the optimal combination of templates and kinematics that best match an observed spectrum, with 
performance tied directly to the resolution and internal consistency of the underlying grids. On the library-construction side, frameworks such 
as \textsc{ProGeny} aim to generate smooth spectral templates by interpolating within and across grids, 
mitigating some of the irregularities inherent in discrete models. However, all interpolation-based approaches face challenges near grid 
edges, where sampling is sparse or physics changes abruptly, and cannot seamlessly merge heterogeneous libraries into a unified continuous 
manifold.

Machine learning offers a promising alternative. Data-driven models that learn mappings between physical parameters and spectra provide differentiable, continuous, and high-speed predictions. Early examples include The Cannon \citep{ness2015cannon}, which introduced a generative framework for label transfer, and The Payne \citep{ting2019payne}, which demonstrated that compact neural networks can emulate Kurucz spectra with high fidelity. Convolutional and deep-learning approaches such as StarNet \citep{fabbro2018application} and \textsc{astroNN} \citep{leung2019deep} show strong performance for stellar label inference, while transformer-based models such as \textsc{GalProTE} \citep{anwar2025galprote} illustrate the flexibility of modern architectures for spectral modelling in noisy environments. Despite these advances, existing ML-based spectral emulators are typically trained on single libraries and thus inherit their parameter-space limits, masked-flux artefacts, and internal inconsistencies. They do not attempt forward spectral synthesis across multiple heterogeneous grids, nor do they treat masked flux as uncertain information to be reconstructed.

Machine learning has also been applied to galaxy-scale spectral analysis and redshift inference, including convolutional models \citep{pasquet2019photometric}, autoencoder frameworks for denoising and dimensionality reduction \citep{portillo2020dimensionality}, and transformer architectures developed for extracting physical properties from low signal-to-noise spectra. These methods highlight the power of ML for modelling complex spectral behaviour, but they primarily address inverse problems rather than forward stellar spectral synthesis.

In this landscape, no existing method learns a unified, continuous spectral manifold spanning multiple high-resolution stellar libraries with overlapping yet heterogeneous parameter ranges. Prior emulators do not address the challenges posed by masked flux regions, library-specific modelling assumptions, or the need for smooth interpolation across inconsistent grid boundaries. Learning across libraries requires reconciling heterogeneous physical assumptions while retaining coherent spectral behaviour. Because each library specialises in different temperature, gravity, and metallicity regimes, a unified model must learn a consistent structure across overlaps and avoid imprinting library-specific artefacts as physical trends. Ensuring smooth, physically plausible predictions in sparsely sampled or rapidly varying regions is essential for applications such as population synthesis, parameter sweeps, and dense template construction.

\begin{figure}[!t]
\centering
\includegraphics[width=1\textwidth]{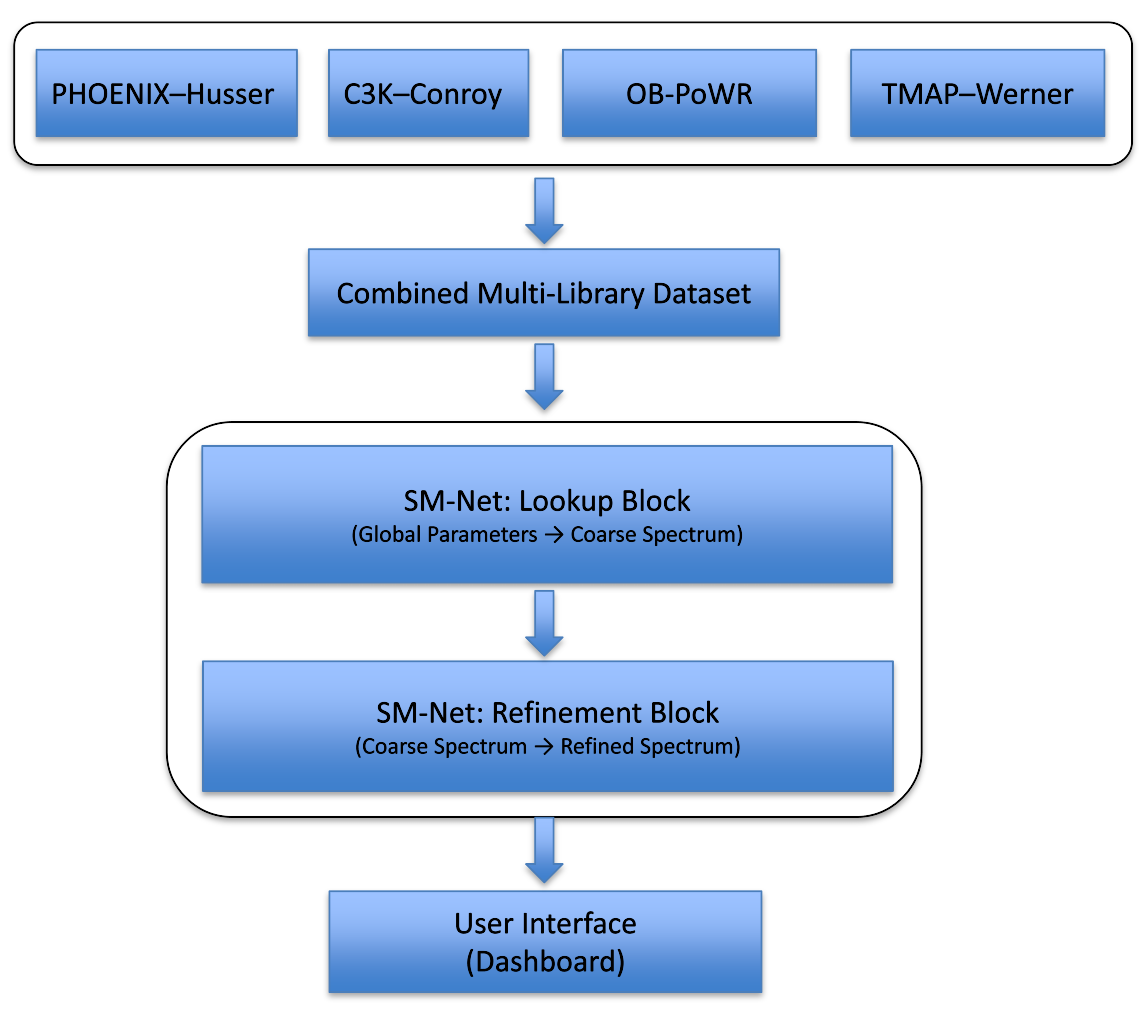}
\caption{
Top-level schematic of the SM-Net model architecture. The four stellar spectral libraries PHOENIX--Husser, C3K--Conroy, OB--PoWR, and TMAP--Werner are first combined into a unified multi-library dataset spanning a broad region of stellar parameter space. SM-Net is then trained on this dataset to generate spectra in two stages: a global parameter-to-coarse-spectrum prediction, followed by a convolutional refinement to produce the final high-fidelity output. The resulting model is accessed through an interactive dashboard that enables real-time spectral generation and exploration.
}
\label{fig:schemetic}
\end{figure}

\begin{table*}[!t]
\centering
\caption{Parameter ranges of the stellar spectrum libraries used in this work. 
N is the number of unique spectra in each library. SM-Net is trained on the combined dataset.}
\label{tab:libraries}
\begin{tabular}{lcccccc}
\hline
\textbf{Library} & \textbf{Type} & \textbf{$T_{\mathrm{eff}}$ ($\mathrm{K}$)} 
& \textbf{$\log g$} & \textbf{$\log Z$} & \boldmath$\lambda$ \textbf{(\AA)} & \textbf{N} \\
\hline
PHOENIX (Husser)      & Base     & 2300--12000   & 0--6        & $-4$--1     & 500--55000         & 7559 \\
C3K (Conroy)          & Base     & 2000--50000   & $-1$--5.5   & $-2.1$--0.5 & 100--$10^{8}$      & 8602 \\
OB (PoWR)             & Hot      & 15000--56000  & 2--4.2      & NA          & 100--$10^{6}$      & 243 \\
TMAP (Werner)         & White    & 20000--190000 & 5--9        & NA          & 3--400000          & 124 \\
\hline
SM-Net (combined)     & ML Model & \textit{2000--190000} 
& \textit{$-1$--9} 
& \textit{$-4$--1} 
& \textit{3000--100000}  
& \textit{$\infty$} \\
\hline
\end{tabular}
\end{table*}

In this work, we present \textbf{SM-Net}, a machine-learning framework designed to learn a continuous spectral manifold from multiple synthetic stellar libraries, as shown in Figure~\ref{fig:schemetic}. By combining the grids of PHOENIX--Husser, C3K--Conroy, OB-PoWR, and TMAP--Werner, we construct a composite dataset, with deterministic handling of exact overlaps, that spans a broader region of stellar parameter space than any individual library. The unified grid covers effective temperatures $T_{\mathrm{eff}} = 2{,}000$--$190{,}000~\mathrm{K}$, surface gravities $\log g = -1$--$9$, and metallicities $\log Z = -4$--$1$, where we use the standard shorthand $\log g \equiv \log_{10}\!\left(g/(\mathrm{cm\,s^{-2}})\right)$ and $\log Z \equiv \log_{10}\!\left(Z/Z_\odot\right)$, with $Z$ denoting the heavy-element mass fraction. SM-Net is trained directly on this combined dataset, treating masked or zero-valued flux samples as unknowns and exploiting local structure in parameter space to infer missing information, over the wavelength range $3{,}000$--$100{,}000~\mathrm{\AA}$. The resulting model can generate spectra at arbitrary locations in parameter space within the joint bounds of the combined libraries (i.e.\ continuous interpolation), achieving inference speeds exceeding $14{,}000$ spectra per second on a single GPU. To support community use, we also provide a browser-based interactive dashboard enabling real-time exploration of the learned manifold.

The remainder of this paper is structured as follows. Section~2 describes the construction of the combined multi-library dataset. Section~3 outlines our preprocessing, normalisation, and augmentation strategy. Section~4 presents the SM-Net architecture and training procedure. Section~5 discusses the cross-library mismatch in detail. Section~6 evaluates performance across the parameter grid. Section 7 presents the web-based interface for SM-Net. Section~8 places SM-Net in the context of existing stellar synthesis workflows, discusses its performance and limitations, and outlines prospects for hybrid physical--machine-learning approaches. Section~9 presents our conclusion and possible future directions.

\section{Dataset Preparation}
\label{dataset}

In this work, we build directly on the homogenised stellar spectral libraries distributed as part of the \textsc{ProGeny} framework, rather than reprocessing the original model grids from their respective primary sources.
These reformatted libraries incorporate several preprocessing steps to improve cross-library consistency and usability, including wavelength resampling onto a common grid, removing problematic spectra, and physically motivated extrapolation of flux at long wavelengths using black-body extensions where required.
Our contribution is therefore not the re-derivation of the underlying stellar atmosphere models, but the unification, augmentation, and modelling of these processed libraries within a single continuous spectral manifold.

The four stellar libraries (PHOENIX–Husser, C3K, OB-PoWR and TMAP) used in this work are summarised in Table \ref{tab:libraries}. The original grids differ in wavelength sampling, parameter coverage, and spectral resolution. In the reformatting performed by \textsc{ProGeny}, all spectra were therefore resampled onto a common wavelength axis spanning $3,000$--$100,000$~\AA. For the Husser PHOENIX spectra, which natively extend only to 55,000~\AA, the missing far-infrared region was extrapolated using a physically motivated blackbody continuation. Because the Husser grid otherwise provides complete photospheric structure up to 55,000~\AA, the blackbody extension contributes only a smooth thermal continuum without adding spurious spectral features. A fifth library, PHOENIX–Allard, was also assessed for possible inclusion; however, its spectra showed highly irregular behaviour at wavelengths below $10{,}000$~\AA\ and for $T_{\mathrm{eff}} < 4{,}000$~$\mathrm{K}$.
Due to these substantial physical inconsistencies, PHOENIX–Allard was excluded from the combined dataset.



\begin{figure*}[!t]
    \centering
    \includegraphics[width=0.75\textwidth]{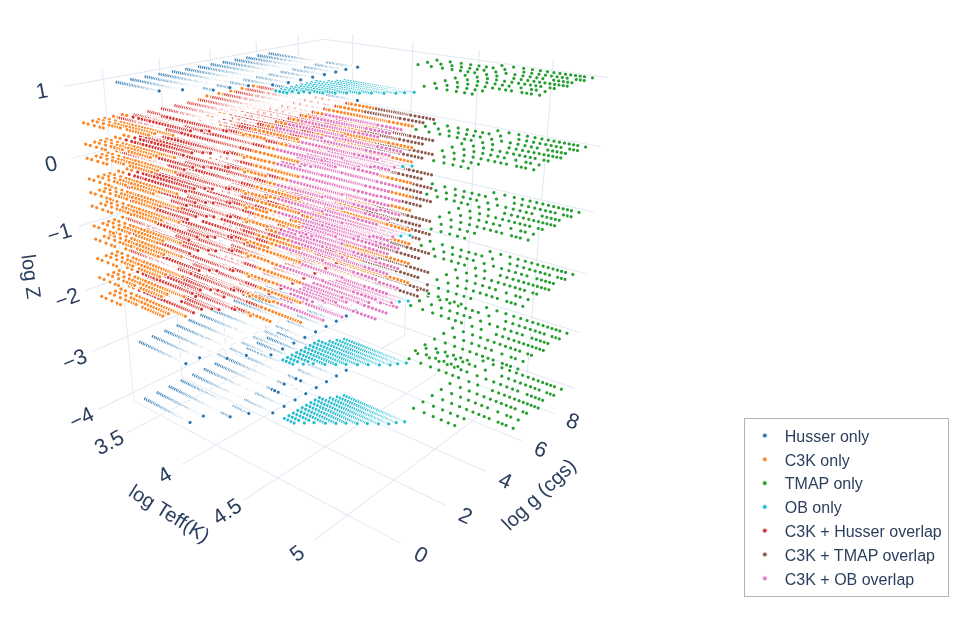}
    \caption{
    A 3D plot representing the coverage of each of the four libraries 
    \textcolor[HTML]{1f77b4}{PHOENIX-Husser}, 
    \textcolor[HTML]{ff7f0e}{C3K},  
    \textcolor[HTML]{17becf}{OB-PoWR}, and 
    \textcolor[HTML]{2ca02c}{TMAP}
    in $(\log_{10} T_{\mathrm{eff}}, \log g, \log Z)$ space. 
    The \textcolor[HTML]{d62728}{overlapping region between C3K and Husser} 
    spans $3.36 \le \log T_{\mathrm{eff}} (K) \le 4.08$, 
    $0 \le \log g \le 5.5$, and $-2.1 \le \log Z \le 0.5$.
    The \textcolor[HTML]{8c564b}{overlapping region between C3K and TMAP} 
    covers $4.30 \le \log T_{\mathrm{eff}} (K) \le 4.70$, 
    $5 \le \log g \le 5.5$, and $-2.1 \le \log Z \le 0.5$.
    The \textcolor[HTML]{e377c2}{overlapping region between C3K and OB-PoWR} 
    covers $4.18 \le \log T_{\mathrm{eff}} (K) \le 4.70$, 
    $2 \le \log g \le 4.5$, and $-2.1 \le \log Z \le 0.5$.
    }
    \label{fig:3DCoverage}
\end{figure*}

\begin{figure*}[!b]
    \centering
    \includegraphics[width=\textwidth]{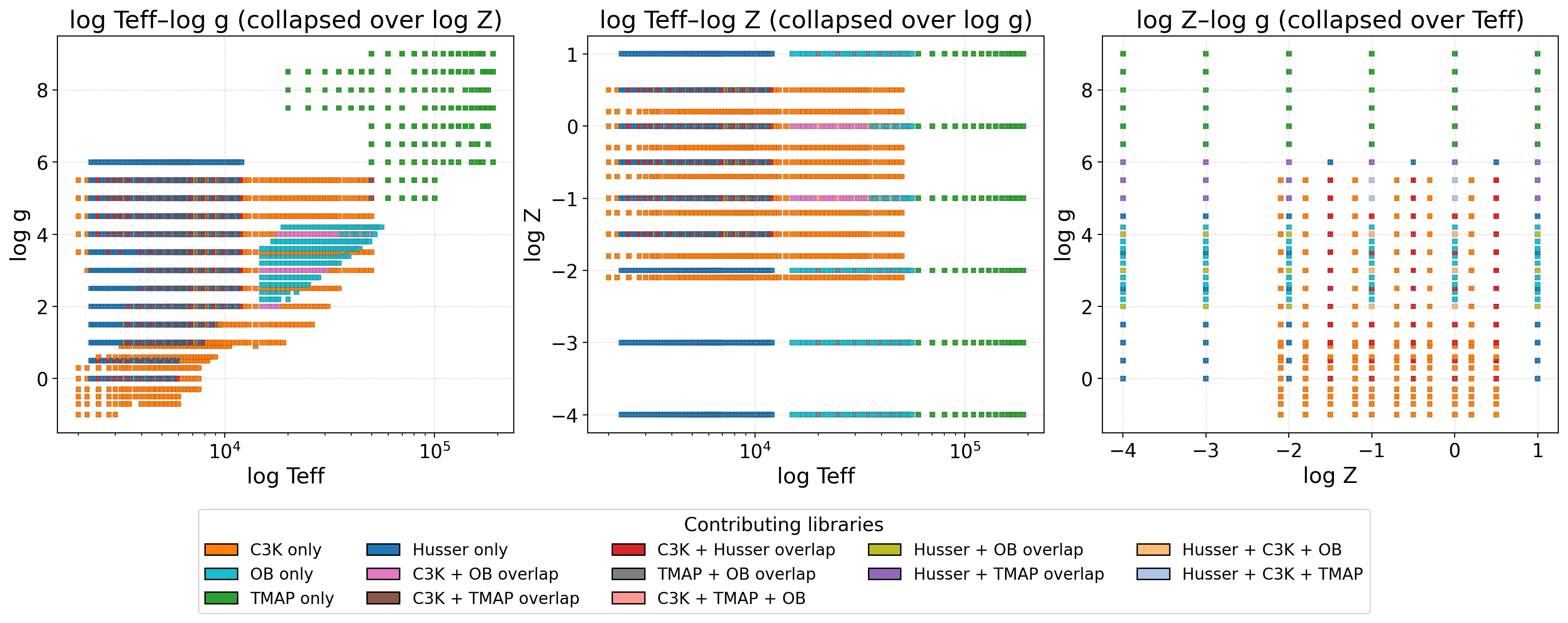}
    \caption{
    Library coverage and overlap on $\log T_{\mathrm{eff}} (K)$ -- $\log g$, $\log T_{\mathrm{eff}} (K)$ -- $\log Z$ and $\log g$ -- $\log Z$ grids.
    }
    \label{fig:lib_overlap_projections}
\end{figure*}

Wavelengths below 3,000~\AA\ were discarded because the four libraries exhibit substantial inconsistencies in this ultraviolet regime, including discontinuities 
and incompatible flux behaviour. The adopted wavelength grid again follows the \textsc{ProGeny} sampling, which is deliberately non-uniform and feature-aware. 
This grid allocates dense sampling in wavelength regions containing strong and rapidly varying spectral features, while permitting much coarser sampling in the 
smooth Rayleigh--Jeans tail at long wavelengths (approaching 100,000~\AA), where the spectrum is nearly featureless. A similar philosophy is adopted in the CB19 stellar population models \citep{plat2019constraints}, which likewise emphasise high fidelity in feature-rich spectral regions.
\textsc{ProGeny} uses the publicly available CB19 stellar spectral grid, which is distributed as a complete model grid rather than described in a dedicated standalone methods paper. The same grid reused in this work therefore provides a practical, well-characterised compromise between spectral fidelity and data volume. After resampling onto this grid, each spectrum contains 11,655 wavelength samples.

The total number of spectra in the four libraries is $16,528$. 
Since \textsc{TMAP} and \textsc{OB-PoWR} do not include metallicity as a model parameter, 
they cannot be directly placed into the three-dimensional $(T_{\mathrm{eff}}, \log g, \log Z)$ 
space used by the other libraries. 
In the grids for these two libraries, metallicity is undefined; all spectra at a given 
$(T_{\mathrm{eff}}, \log g)$ are identical regardless of any value of $\log Z$ that might be assigned. 
To embed \textsc{TMAP} and \textsc{OB-PoWR} into a unified three-parameter space without 
altering their data structure, each spectrum from these libraries was replicated six times 
with assigned metallicities $\log Z \in \{-4,-3,-2,-1,0,1\}$. 
This covers the entire metallicity range on the grid, with a 1 dex interval. 
This procedure enables dimensional compatibility with the other libraries, allowing the 
network to learn the intended \textit{grid behaviour}: in the region of parameter space 
covered exclusively by \textsc{TMAP} or \textsc{OB-PoWR}, the spectra are invariant under 
changes in $\log Z$. 
We emphasise that this invariance is a limitation of the underlying training data 
rather than a physical property of the stars. 
While hot stellar spectra are physically affected by metallicity, these effects are absent 
in the specific source grids used here for this parameter regime.


After replicating the spectra for \textsc{TMAP} and \textsc{OB-PoWR} 6 times, the total spectra for these libraries increase to 744 (124$\times$6) and 1,458 (243$\times$6), respectively. This increased the total number of spectra for all 4 libraries to 18,368. But in regions where $(T_{\mathrm{eff}}, \log g, \log Z)$ parameters coincided exactly, only one spectrum was retained following a deterministic priority order: \textsc{TMAP} $>$  \textsc{PHOENIX}--Husser $>$ \textsc{OB-PoWR} $>$ \textsc{C3K}. This ordering reflects the fact that \textsc{TMAP}, \textsc{PHOENIX}--Husser, and \textsc{OB-PoWR} each specialise in distinct regions of parameter space, while \textsc{C3K} both provides complementary coverage in overlapping regions and contributes unique coverage at lower effective temperatures where no other library is available. As a result, all $927$ discarded spectra belong to \textsc{C3K} and arise exclusively from regions of exact parameter overlap. This step removes physically distinct spectra that share identical labels and enforces a consistent reference spectrum across overlapping regions of the grid. The final combined dataset contains $17,441$ spectra. 

This construction step is important for interpreting the role of SM-Net. The model is trained to emulate the final composite dataset defined by the preprocessing and priority rules above, rather than to infer a physically unique reconciliation between libraries when multiple source spectra share identical labels. In exact-overlap regions, SM-Net therefore learns the retained reference spectrum associated with the imposed priority ordering.

The 3D parameter-space coverage of the combined dataset is shown in Figure \ref{fig:3DCoverage}, which illustrates that the union of the four libraries spans a much broader and more heterogeneous region of $(T_{\mathrm{eff}}, \log g, \log Z)$ than any single library alone. The variable densities for four libraries in different regions, as well as the irregular shape of the combined grid is also prominent. Figures \ref{fig:lib_overlap_projections} represent the 2D projections of the 3D grid in 3 different planes for further clarity on the coverage, and the overlap for 2 parameters per figure. The $T_{\mathrm{eff}}$ axis in these projections is on a logarithmic scale.

To minimise overlap between the training and validation sets, we constructed the data splits directly in parameter space rather than by random sampling.
For each of the three parameters $(T_{\mathrm{eff}}, \log g, \log Z)$, the unique sorted values were assigned to the training and validation sets following a fixed repeating pattern, in which two consecutive values were allocated to the training set and the next two to the validation set.
This procedure was applied independently along each parameter axis, such that a spectrum assigned to the training set does not share any of its parameter values with any spectrum in the validation set.
The only deliberate exception occurs at the grid boundaries, where the extreme values (the first two and last two grid points along each axis) are included in both splits to ensure stable behaviour near the edges and corners of the parameter space.
Here, boundary cases refer specifically to spectra lying at the extremes of the sampled ranges in $T_{\mathrm{eff}}$, $\log g$, or $\log Z$, and are shared solely to anchor interpolation at these limits.
Due to the irregular shapes and incomplete overlaps among the individual library grids, only 17 spectra are shared between the training and validation sets.



Let's consider 3 sorted parameter sets from the combined dataset
\[
\begin{aligned}
T_{\mathrm{eff}} &= \{t_1,t_2,t_3,t_4,t_5,...\},\\
\log g &= \{g_1,g_2,g_3,g_4,g_5,...\},\\
\log Z &= \{z_1,z_2,z_3,z_4,z_5,...\}.
\end{aligned}
\]

With 2 boundary values (subscript $_1$ and $_2$) shared between train and validation, and applying the 2-train and 2-validation pattern independently to each axis, gives

{\small
\[
\begin{aligned}
T_{\mathrm{eff}}^{\mathrm{train}} &= \{t_1,t_2,t_3,t_4,t_7,...\}, &\quad
T_{\mathrm{eff}}^{\mathrm{val}} &= \{t_1,t_2,t_5,t_6,...\},\\
\log g^{\mathrm{train}} &= \{g_1,g_2,g_3,g_4,g_7,...\}, &\quad
\log g^{\mathrm{val}} &= \{g_1,g_2,g_5,g_6,...\},\\
\log Z^{\mathrm{train}} &= \{z_1,z_2,z_3,z_4,z_7,...\}, &\quad
\log Z^{\mathrm{val}} &= \{z_1,z_2,z_5,z_6,...\}.
\end{aligned}
\]
}

\begin{table}[!t]
\centering
\caption{Illustrative assignment of spectra to train/validation/test sets based on the 2--2 parameter split pattern. 
Ticks indicate membership; boundary cases appear in both the training and validation sets.}
\label{tab:split_example}
\begin{tabular}{ccc|ccc}
\hline
$T_{\mathrm{eff}}$ & $\log g$ & $\log Z$ & Train & Val & Test \\
\hline
$t_1$ & $g_1$ & $z_1$ & \cmark & \cmark &       \\  
$t_2$ & $g_1$ & $z_2$ & \cmark & \cmark &       \\  
$t_2$ & $g_3$ & $z_3$ & \cmark &        &        \\  
$t_3$ & $g_2$ & $z_5$ &        &        & \cmark \\ 
$t_5$ & $g_3$ & $z_1$ &        &        & \cmark \\ 
$t_5$ & $g_1$ & $z_3$ &        &        & \cmark       \\  
$t_5$ & $g_5$ & $z_5$ &        &  \cmark &       \\  %
$t_1$ & $g_3$ & $z_6$ &        &  &  \cmark      \\  %
\vdots & \vdots & \vdots & \vdots & \vdots & \vdots \\
\hline
\end{tabular}
\end{table}

A spectrum belongs to the training set only if \emph{all three} of its parameter values fall inside the training subsets above; similarly for the validation set. Some random example combinations are shown in Table \ref{tab:split_example}. Any spectrum whose parameters do not fall entirely within either training or validation subsets is assigned to the test set. This demonstrates that the split is defined in \emph{parameter space}, not by random splits, and that the resulting partitions reflect the true structure of the irregular multi-library grid.

Because the combined grid is formed from four libraries with highly irregular and different parameter coverage, the vast majority of possible $(T_{\mathrm{eff}}, \log g, \log Z)$ combinations do not correspond to any real spectrum in any library. Such combinations are discarded and therefore are not allocated to any split.

This splitting scheme forces the model to validate its results on parameters never encountered during training, hence the validation set provides a strict measure of generalisation for early stopping. After assigning $3,538$ samples to the training set and $2,386$ to the validation set, the test set comprises the remaining $11,530$ spectra not in either subset. Although this produces a relatively large test split and smaller train/validation splits, which is unconventional in machine learning, the resulting parameter distributions remain well spread across the grid, ensuring adequate training coverage and fair evaluation of interpolation capability at unseen parameter combinations.

\begin{figure}[!b]
    \centering
    \includegraphics[width=1\textwidth]{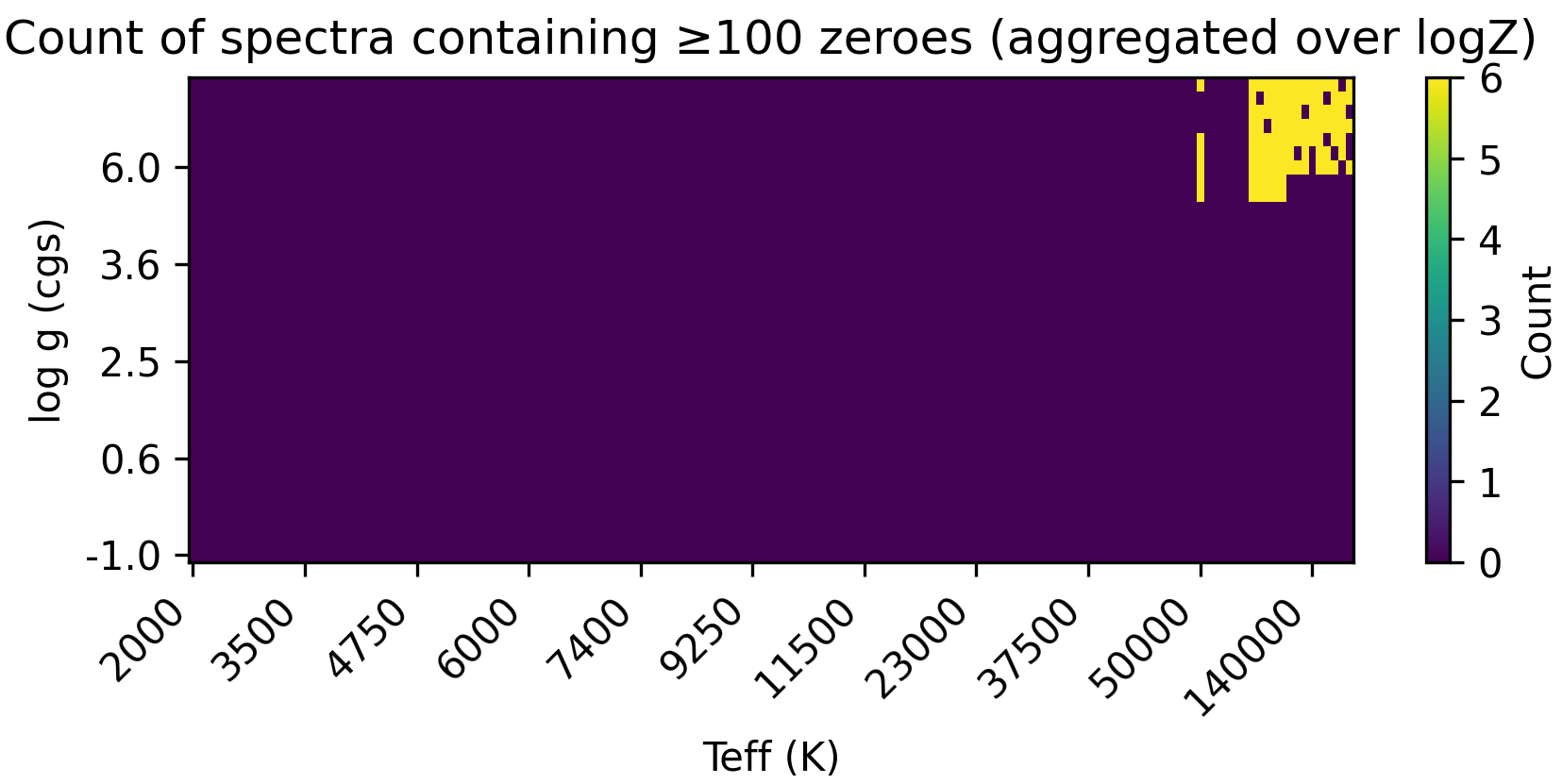}
    \caption{
    Distribution of zero/unknown flux values across the dataset of combined grids. Counts are aggregated over logZ values.
    }
    \label{fig:ZeroFlux}
\end{figure}

Missing or invalid flux values were identified through NaN and zero-value detection, with zero-valued regions treated as masked intervals rather than physical flux. A bounded zero-run analysis was applied to detect internal masked regions without trimming spectra, preserving as much surrounding information as possible. Figure \ref{fig:ZeroFlux} shows the distribution of unknown flux values in the combined dataset, where more than 100 flux values per spectra were missing. For the wavelength range of $3,000$ to $100,000$~\AA, all of the spectra with missing flux regions belong to the TMAP library.

For model training and evaluation, flux values in the combined dataset were mapped to the interval $[0,1]$ using a log1p-based scaling.
This transformation compresses the large dynamic range of stellar spectra while remaining strictly monotonic in the original flux.
In practice, we adopt a numerically stable implementation in which zero-valued flux elements remain exactly zero, and the logarithmic scale is set adaptively using a small percentile of the positive flux distribution to avoid sensitivity to extreme outliers.
The resulting spectra are normalised using a dataset-wide maximum flux, ensuring consistent scaling across libraries with differing absolute normalisations.
Although this nonlinear transformation alters linear flux ratios, it preserves the relative ordering of spectral features and stabilises optimisation without introducing spurious structure into masked or low-flux regions.

\begin{table}[!b]
    \centering
    \caption{Number of spectra in each dataset split before and after augmentation based on spectral uniqueness.}
    \label{tab:aug_sizes}
    \begin{tabular}{lrrr}
        \hline
        Split & Original $N$ & Augmented $N$ & Increase factor \\
        \hline
        Train       & 3,538  & 5,784  & 1.63 \\
        Validation  & 2,386  & 3,845  & 1.61 \\
        Test        & 11,530 & N/A & N/A \\
        \hline
    \end{tabular}
\end{table}

\section{Spectral Similarity and Augmentation Strategy}

A key challenge in training an ML model is constructing a balanced training dataset. Unlike typical classification or regression problems where class imbalance arises from having many examples of some classes and few of others, this combined dataset contains a single spectrum for each unique triplet/class of $(T_{\mathrm{eff}}, \log g, \log Z)$. In this case, the imbalance instead arises from the underlying structure of the stellar parameter grid. Certain parts of the grid are densely sampled and contain spectra whose shapes change only gradually across neighbouring points, resulting in many regions with nearly identical spectra. In contrast, other areas of the grid exhibit sharp spectral transitions but contain relatively few grid points. This leads to a biased distribution of spectral shapes, with some spectral morphologies appearing abundantly while others are represented by only a handful of examples.

\begin{figure}[!t]
    \centering
    \includegraphics[width=1\textwidth]{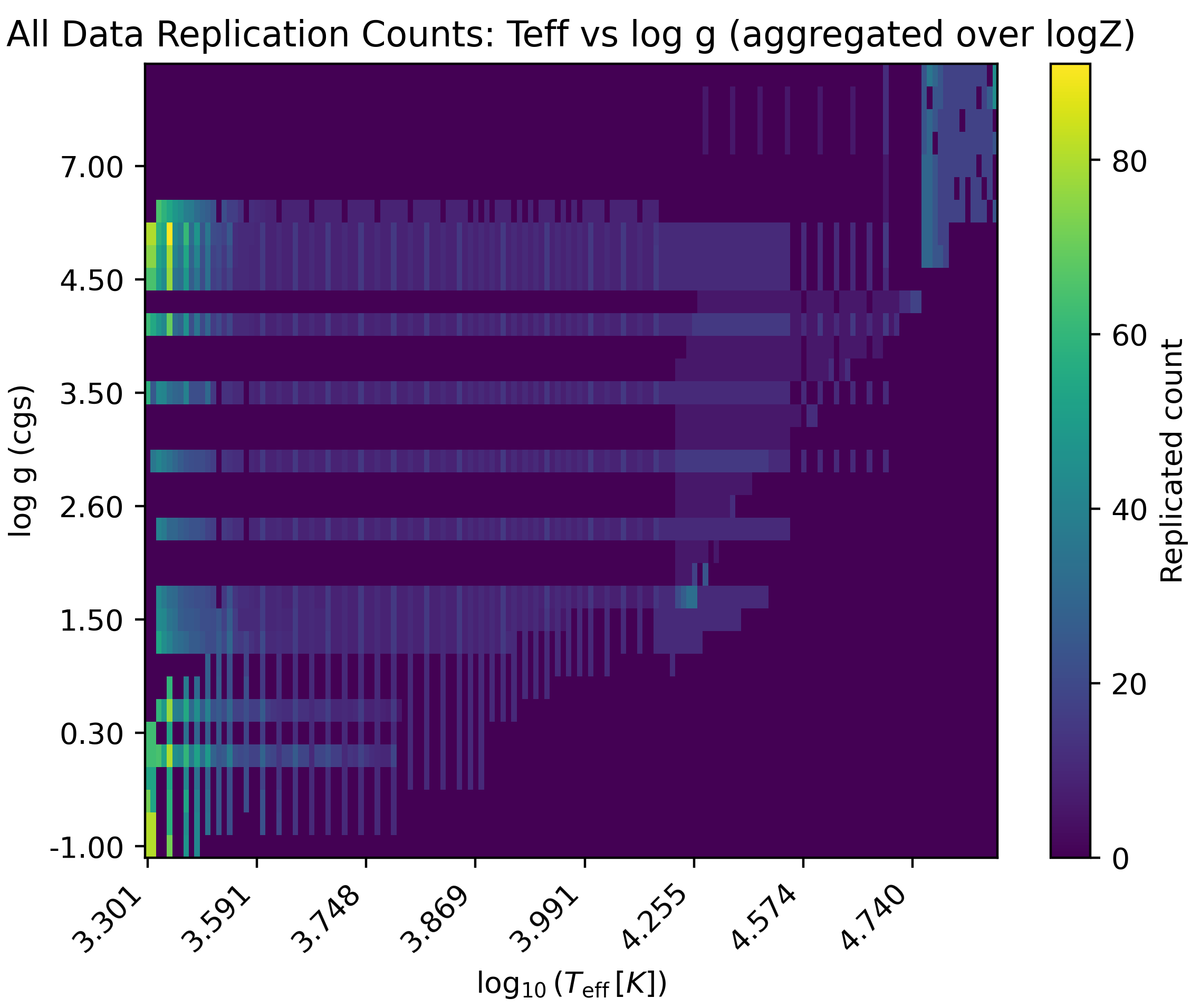}
    \caption{
    The replication counts for the entire dataset based on spectral dissimilarity, aggregated over $\log Z$. 
    }
    \label{fig:replication}
\end{figure}

To correct for this imbalance, we computed pairwise cosine similarity across the combined dataset, supplemented by a masked L1 distance that excludes wavelength intervals with zero flux. These similarities quantify the uniqueness of each spectrum in a physically meaningful way, emphasising real morphological differences. Spectra that were more distinctive were assigned higher augmentation counts, up to a maximum of 10, based on dissimilarity level, while highly similar spectra received no augmentation. As a result, about 20\% of the spectra in the combined dataset are repeated more than once, whereas the remaining 80\% appear only once. The final numbers of spectra in the splits after augmentation are shown in Table~\ref{tab:aug_sizes}, and Figure \ref{fig:replication} shows a graphic representation of the replication counts in the $T_{\mathrm{eff}}$ and $\log g$ grid, aggregated over $\log Z$. The bulk of the spectra with distinct shapes lie in the low-temperature range, and their high replication counts are indicated by yellow. In some regions, such as $\log g = 2$ and $T_{\mathrm{eff}} = 18,000$, the overlap in libraries across logZ also contributes to the higher count of spectra.

The validation set was also augmented because early stopping relies on its loss as a proxy for generalisation, and therefore it must reflect the full diversity of spectral morphologies, including rare cases, rather than only the most common spectral shapes. Augmentation of the validation set ensures that early stopping is triggered by robust performance across all spectral shapes, whereas the spectra in the test set were not replicated, so that the test set remains a strictly unbiased measure of the model's accuracy. Overall, the augmentation helps rare spectral shapes exert sufficient influence during training and validation, and prevents the model from overfitting to the most common spectral shapes in the dataset.

\begin{figure*}[!t]
    \centering
    \includegraphics[width=1\textwidth]{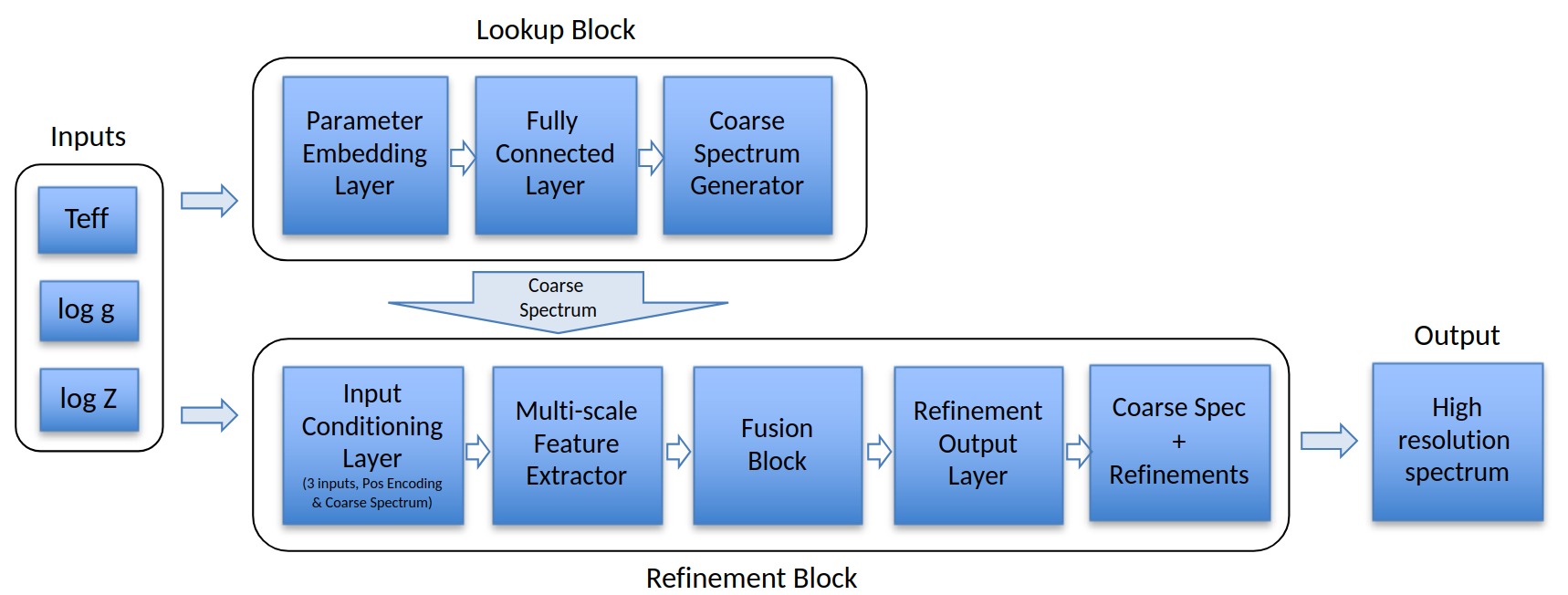}
    \caption{
    Block-level architecture of SM-Net, showing the two-stage mapping from stellar parameters to a high-resolution synthetic spectrum. The Lookup Block takes the input parameters (effective temperature, surface gravity, and metallicity) and produces a coarse, low-frequency spectral estimate through a sequence of embedding, fully connected, and projection layers. This coarse spectrum, together with the physical parameters and positional encodings, is passed to the Refinement Block, where convolutional feature extractors and residual convolutional blocks learn wavelength-dependent corrections. The refinement output is added to the coarse spectrum to produce the final high-resolution stellar spectrum. 
    }
    \label{fig:arch}
\end{figure*}

\section{Model Architecture and Training}
\label{ML_model}
SM-Net consists of two main blocks: a global parameter--to--spectrum predictor (\emph{Lookup Block}) and a convolutional refinement module (\emph{Refinement Block}), as shown in Figure~\ref{fig:arch}. Together, these blocks map the three input stellar parameters
\[
\mathbf{x} =
\begin{bmatrix}
T_{\mathrm{eff}} & \log g & \log Z
\end{bmatrix},
\]
to a full-resolution spectrum of length $L = 11,655$. The Lookup Block provides a smooth coarse estimate, while the Refinement Block adds wavelength-dependent corrections to produce the final high-fidelity output.

\subsection{Lookup Block: Global Parameter-to-Spectrum Mapping}

The first stage of SM-Net is the \emph{Lookup Block}, which transforms the three-dimensional input parameters into a coarse, low-frequency spectral prediction. In terms of the architecture in Figure~\ref{fig:arch}, this stage consists of:
\begin{itemize}
\item a \textbf{Parameter Embedding Layer} that expands the physical inputs into a higher-dimensional representation,
\item three \textbf{Fully Connected Layers}, with dimensions 
$3 \rightarrow 4,096 \rightarrow 16,384 \rightarrow 11,655$, 
implemented using residual MLP blocks,
\item a \textbf{Coarse Spectrum Generator} that produces an initial spectral estimate.
\end{itemize}

The embedded parameters pass through two residual fully connected layers:
\begin{align}
\mathbf{h}_1 &= f_1(\mathbf{x}) + P_1(\mathbf{x}), \label{eq:h1}\\[4pt]
\mathbf{h}_2 &= f_2(\mathbf{h}_1) + P_2(\mathbf{x}), \label{eq:h2}
\end{align}
where $f_1$ and $f_2$ are residual MLP blocks, and $P_1, P_2$ are linear skip projections that forward the physical parameters directly to deeper layers.

The \textbf{Coarse Spectrum Generator} produces the initial coarse spectrum:
\begin{equation}
\mathbf{s}_{\mathrm{coarse}} = W_3 \mathbf{h}_2 + P_3(\mathbf{x}),
\label{eq:coarse}
\end{equation}
where $W_3$ is a learned linear operator and $P_3$ is the final linear layer of this block,
mapping output to the spectral dimension.


A learnable per-wavelength baseline vector $\mathbf{b} \in \mathbb{R}^{L}$ is initialised to the median flux of the training data. This baseline acts as a stabilising prior on the predicted spectrum. It is applied using a Softplus transformation,
\begin{equation}
\mathbf{s}_0 = \mathrm{Softplus}_\beta\!\left( \mathbf{s}_{\mathrm{coarse}} + \mathbf{b} \right),
\label{eq:baseline}
\end{equation}
which guarantees non-negative output fluxes. The parameter $\beta$ controls the sharpness of the Softplus activation.

\subsection{Refinement Block: Residual Spectral Refinement}

The second stage of SM-Net is the \emph{Refinement Block}, which applies a series of convolutional operations to add detailed wavelength-dependent corrections to the coarse prediction. Architecturally, this block consists of:
\begin{itemize}
    \item an \textbf{Input Conditioning Layer} that integrates the learned conditioning vector, a fixed three-channel positional encoding, and the coarse spectrum;
    \item a \textbf{Multi-scale Feature Extractor} consisting of three parallel convolutional branches with kernel sizes of 7, 15, and 23;
    \item a \textbf{Fusion Block} that aggregates features from the parallel branches;
    \item a \textbf{Refinement Output Layer} that predicts the additive spectral correction.
\end{itemize}

The stellar parameters are first expanded along the wavelength axis by a learned linear map:
\begin{equation}
    \mathbf{c} = Q\,\mathbf{x} + \mathbf{b},
    \label{eq:condproj}
\end{equation}
where
\begin{itemize}
    \item $\mathbf{x} \in \mathbb{R}^{D}$ is the input parameter vector (where $D = 3$);
    \item $Q \in \mathbb{R}^{L \times D}$ is the learned weight matrix projecting parameters onto the spectral axis;
    \item $\mathbf{b} \in \mathbb{R}^{L}$ is the bias vector;
    \item $\mathbf{c} \in \mathbb{R}^{L}$ is the resulting conditioning vector.
\end{itemize}
Here $L$ denotes the length of the spectrum.

To provide the refinement module with an explicit notion of spectral position, we generate a fixed three-channel positional encoding $\mathbf{P} \in \mathbb{R}^{3 \times L}$. For a normalized wavelength index $t_i \in [0, 1]$ at bin $i$, the encoding channels are defined as:
\begin{equation}
    \mathbf{P}_i = 
    \begin{bmatrix}
    t_i \\
    \sin(2\pi t_i) \\
    \cos(2\pi t_i)
    \end{bmatrix}.
    \label{eq:encoding}
\end{equation}

The \textbf{Input Conditioning Layer} stacks the coarse spectrum $\mathbf{s}_0$, the conditioning vector $\mathbf{c}$, and the positional encoding $\mathbf{P}$:

\begin{equation}
    \mathbf{X} = 
    \text{stack}(
    \mathbf{s}_0,\,
    \mathbf{c}
    ) \Rightarrow \mathbf{X} \in \mathbb{R}^{5 \times L}.
    \label{eq:Xstack}
\end{equation}

Note that $\mathbf{s}_0$ and $\mathbf{c}$ are stacked (dim=1) and then concatenated with $\mathbf{P}$.

The \textbf{Multi-scale Feature Extractor} processes $\mathbf{X}$ through three parallel branches to capture features at different scales. For each kernel size $k \in \{7, 15, 23\}$, the output $\mathbf{u}_k$ is computed as:
\begin{equation}
    \mathbf{u}_k = \mathcal{D}\left( \phi\left( \mathrm{Conv}_k\left( \mathrm{pad}_k(\mathbf{X}) \right) \right) \right),
    \label{eq:multiscale}
\end{equation}
where $\mathrm{pad}_k(\cdot)$ denotes replicate padding, $\phi$ is the GELU activation, and $\mathcal{D}$ denotes Dropout.
The feature maps from the parallel branches, each corresponding to a different kernel size $k$, are concatenated along the channel dimension to form a composite multi-scale feature map $\mathbf{U}$. This aggregation allows the network to process local spectral lines and broader continuum features simultaneously. The composite map is processed by a \textbf{Fusion Block}:

\begin{equation}
    \mathbf{h} = \mathcal{D}\left( \phi\left( \mathrm{Conv}_{\mathrm{fuse}}(\mathbf{U}) \right) \right).
    \label{eq:fuse1}
\end{equation}

Here, $\mathrm{Conv}_{\mathrm{fuse}}$ represents a convolutional layer designed to mix information across the spectral channels. Finally, the \textbf{Refinement Output Layer} projects these fused features down to a single dimension to produce the scalar spectral correction $\Delta\mathbf{s}$:

\begin{equation}
    \Delta\mathbf{s} = \mathrm{Conv}_{\mathrm{out}}(\mathbf{h}).
    \label{eq:fuse2}
\end{equation}

The final high-resolution spectrum is obtained via a residual connection, where the learned correction is added to the initial coarse estimate:

\begin{equation}
    \mathbf{s}_{\mathrm{final}} = \mathbf{s}_0 + \Delta\mathbf{s}.
    \label{eq:sfinal}
\end{equation}

\subsection{Loss Function and Treatment of Masked Flux}

To handle masked or zero-valued flux (originating from incomplete line lists or numerical issues), SM-Net uses a weighted mean-squared error (MSE) loss:
\begin{equation}
\mathcal{L} =
\frac{1}{BL}
\sum_{i=1}^{B}
\sum_{\lambda=1}^{L}
w_\lambda\, m_{i,\lambda}\,
\left( s_{i,\lambda} - t_{i,\lambda} \right)^2,
\label{eq:loss}
\end{equation}
where:
\begin{itemize}
\item $B$ is the batch size,
\item $w_\lambda$ is a per-wavelength weight,
\item $t_{i,\lambda}$ and $s_{i,\lambda}$ are the target and predicted fluxes,
\item $m_{i,\lambda}$ is a binary mask:
\[
m_{i,\lambda} =
\begin{cases}
0, & t_{i,\lambda} = 0 \ (\text{masked or uncertain flux}),\\
1, & \text{otherwise}.
\end{cases}
\]
\end{itemize}

Masked flux values therefore exert no penalty, allowing the model to reconstruct missing structure by relying on neighbouring wavelengths and on the broader learned relationships encoded in the network.


\subsection{Training Strategy}

\begin{figure}[!t]
    \centering
    \includegraphics[width=1\textwidth]{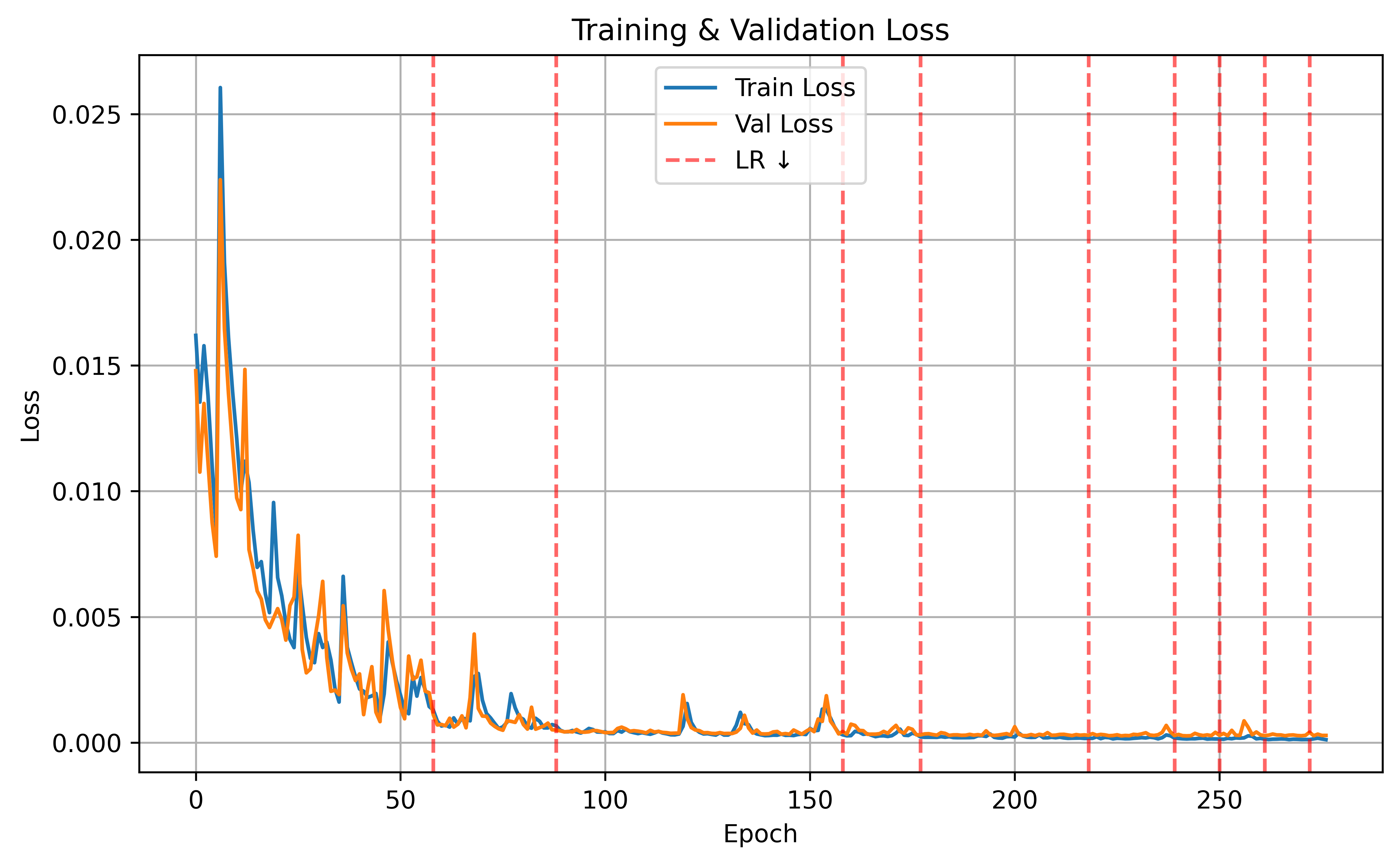}
    \caption{
    Training and validation losses throughout the training process. Red lines indicate the epochs at which the learning rate was reduced and the batch size was reduced, both by a factor of 0.8.
    }
    \label{fig:loss}
\end{figure}

SM-Net model was first trained, evaluated and refined separately on each of the four core libraries.
This initial evaluation phase not only guided model optimisation but also revealed regions within individual libraries where the spectra exhibited internal inconsistencies. Wavelengths below 3,000~\AA\ were removed as a result of inconsistencies highlighted during these individual trainings. 

Model training used the Adam optimiser for Learning Rate (LR) scheduling, with an initial LR of $4\times 10^{-4}$ and an initial batch size of $1024$.  
Whenever the validation loss failed to improve for ten consecutive epochs, both the LR and batch size were reduced by a factor of $0.8$, with a minimum batch size capped at $128$. Early stopping was triggered after $50$ epochs without improvement. This decay schedule in LR and reduction in training batch size, shown as red dotted lines in Figure~\ref{fig:loss}, served two complementary roles. Large batches in early epochs stabilised the optimisation landscape and enabled the network to learn broad, global spectral trends, whereas progressively smaller batches introduced beneficial gradient noise that helped the optimiser escape shallow minima and focus on fine-grained spectral structure. Reducing the LR in tandem ensured that later stages of training refined subtle features without overshooting, which is essential for high-resolution spectra, where neighbouring wavelengths differ only marginally in flux. Empirically, this coupled decay of batch size and LR yielded smoother convergence than adjusting either hyper-parameter alone.

We enabled automatic mixed-precision with CUDA, using \texttt{bfloat16} throughout training with no gradient scaling required. This reduced memory pressure during convolutional refinement without degrading accuracy. Training required a peak GPU memory of $21$\,GB, whereas inference with a batch size of 256 spectra requires about $4.5$\,GB of GPU memory.  
The model achieves rates of approximately $14,000$ spectra per second on a single NVIDIA RTX 4090 GPU. For maximum throughput and full GPU utilisation, the number of inferred spectra should exceed a few thousand. For small inference jobs, the GPU is under-utilised, kernel-launch overheads dominate, and the resulting throughput is noticeably lower than the peak rate.

\section{Cross-library physics mismatches}

\begin{figure}[!b]
    \centering
    \includegraphics[width=1\textwidth]{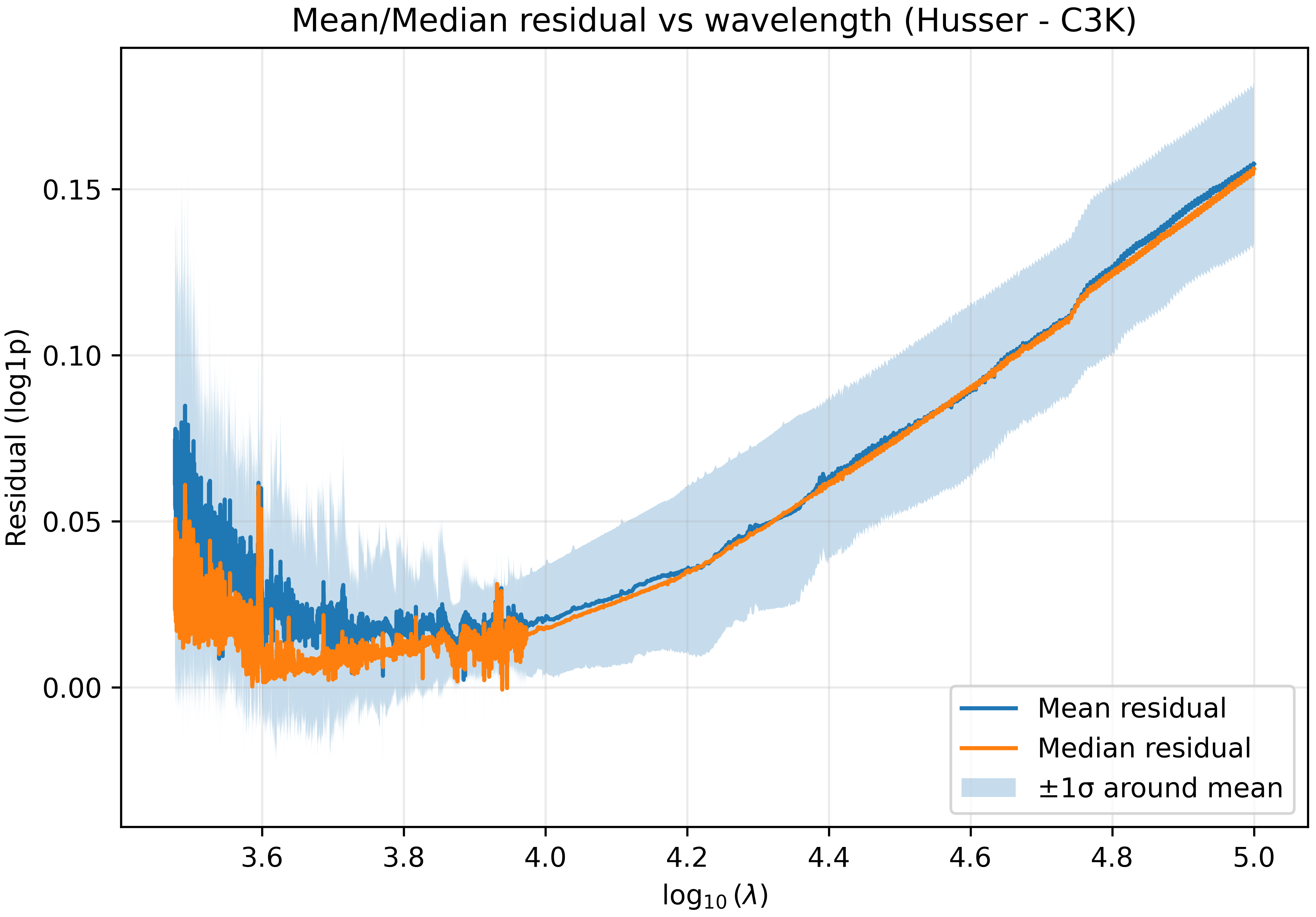}
    \caption{
    Mean and median residual of spectra between Husser and C3K libraries as a function of $\log_{10}(\lambda)$, across 783 matched $(T_{\mathrm{eff}}, \log g, \log Z)$ points on the grid between two libraries. The shaded region indicates the $\pm1\sigma$ dispersion of the residuals.
    }
    \label{fig:residuals}
\end{figure}

\begin{figure}[!t]
    \centering
    \includegraphics[width=1\textwidth]{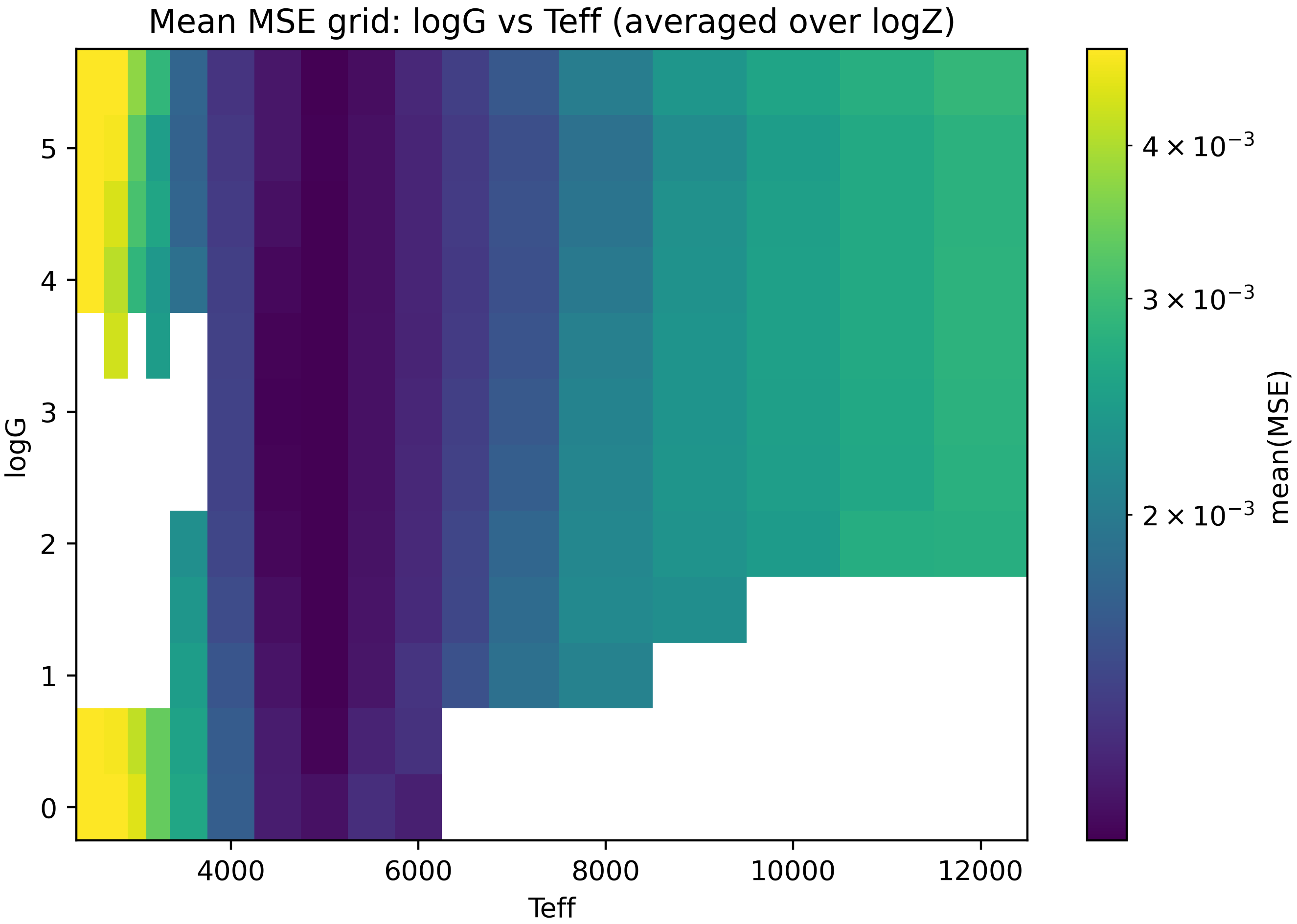}
    \caption{
    Mean of Mean Square Difference (Teff vs logG) between the common spectra of Husser and C3K libraries. The mean is computed over the logZ axis.
    }
    \label{fig:MSE_C3K_Huss}
\end{figure}

\begin{figure}[!b]
    \centering
    \includegraphics[width=1\textwidth]{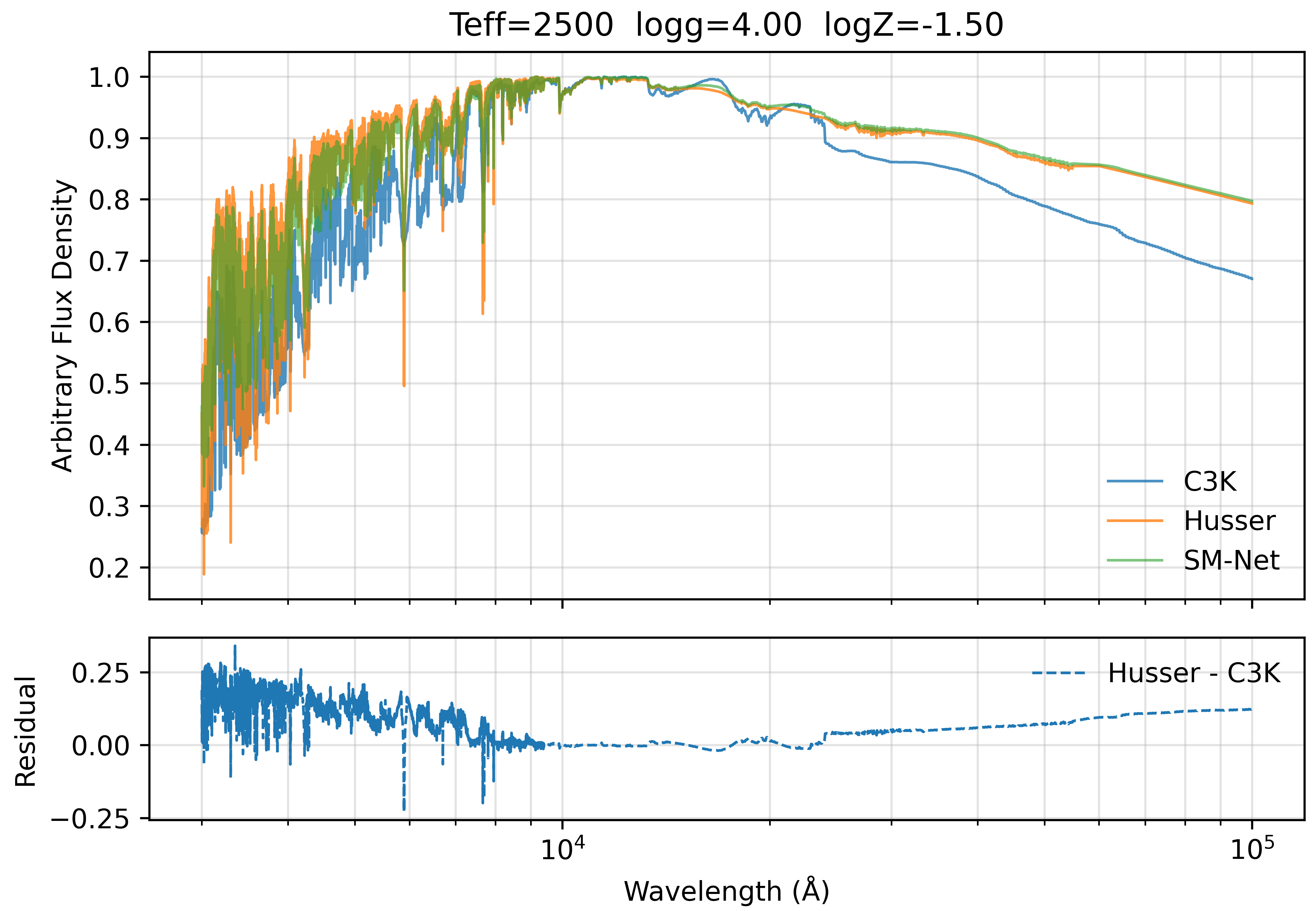}
    \caption{
    C3K and PHOENIX-Husser spectrum for $T_{\mathrm{eff}} = 2,500~\mathrm{K}$, $\log g = 4.0$, and $\log Z = -1.5$, and the SM-Net generated spectrum for the same parameters.
    The residual highlights the underlying differences in the physics between the two libraries.
    }
    \label{fig:mismatch}
\end{figure}

The mismatch in the underlying physics between different stellar libraries directly affects the accuracy that SM-Net can achieve. This mismatch between some libraries is significant. Figure~\ref{fig:residuals} shows the mean and median residuals when comparing the 783 spectra with the exact same $T_{\mathrm{eff}}$, $\log g$, and $\log Z$ labels between the PHOENIX–Husser and C3K libraries, highlighting the mismatch in the physics between the two libraries. Figure \ref{fig:MSE_C3K_Huss} shows the 2D Teff-logG grid of the mean MSE (difference) between the common spectra of the two libraries, averaged on the logZ axis. The lower temperatures exhibit the largest physics mismatch between the libraries, as indicated by the bright yellow colour.

Figure~\ref{fig:mismatch} illustrates the mismatch using two original spectra with identical parameters ($T_{\mathrm{eff}}=2,500$~$\mathrm{K}$, $\log g=4.0$, $\log Z=-1.5$) drawn from the two libraries, alongside their residual spectrum and the output produced by SM-Net for the same parameters. All three spectra are normalised to a peak flux of unity to enable direct comparison. At shorter wavelengths, the flux difference between the spectra from two libraries exceeds 20\% and remains as high as $\sim10\%$ at the longest wavelengths, highlighting a substantial underlying physical mismatch. The SM-Net reconstruction more closely follows the PHOENIX–Husser spectrum, consistent with the dataset design (Section~\ref{dataset}), where PHOENIX–Husser was preferred over C3K whenever both libraries provided spectra for the same parameters. Consequently, in regions where PHOENIX–Husser overlaps with C3K, SM-Net naturally aligns with PHOENIX–Husser that was preferred during dataset construction. 

The consequences of these cross-library mismatches become evident in the reconstruction-error distributions presented in the next section, particularly near overlap boundaries and in sparsely sampled regimes.

\section{Evaluation and Performance}

\begin{table*}[!t]
    \centering
    \caption{Summary of SM-Net reconstruction quality across the combined multi-library dataset. 
    For each split, $N$ is the number of spectra, mean $\mathrm{MSE}$ is the mean per-spectrum 
    mean squared error, and $\mathrm{RMS}_\mathrm{flux}$ is the corresponding 
    root-mean-square flux deviation. Percentiles and min/max values are computed over the 
    per-spectrum MSE distribution.}
    \label{tab:smnet_quality}
    \begin{tabular}{lrrrrrrrr}
        \hline
        Split & $N$ & mean $\mathrm{MSE}$ 
              & $\mathrm{RMS}_\mathrm{flux}$ [\%] 
              & Median MSE 
              & 10th perc. MSE 
              & 90th perc. MSE
              & Min MSE
              & Max MSE \\
        \hline
        Train & 3,538 
              & $1.47\times10^{-5}$ 
              & $0.38$ 
              & $4.20\times10^{-6}$ 
              & $1.13\times10^{-6}$ 
              & $3.17\times10^{-5}$
              & $1.78\times10^{-7}$ 
              & $6.34\times10^{-4}$ \\
        Validation & 2,386 
              & $2.62\times10^{-5}$ 
              & $0.51$ 
              & $5.59\times10^{-6}$ 
              & $1.56\times10^{-6}$ 
              & $5.35\times10^{-5}$  
              & $2.59\times10^{-7}$ 
              & $9.81\times10^{-4}$ \\
        Test & 11,530 
              & $2.34\times10^{-5}$ 
              & $0.48$ 
              & $5.09\times10^{-6}$ 
              & $1.39\times10^{-6}$ 
              & $4.23\times10^{-5}$ 
              & $1.86\times10^{-7}$ 
              & $3.73\times10^{-3}$ \\
        \hline
    \end{tabular}
\end{table*}

\begin{figure}[!b]
    \centering

    \begin{subfigure}{1\textwidth}
        \centering
        \includegraphics[width=\textwidth]{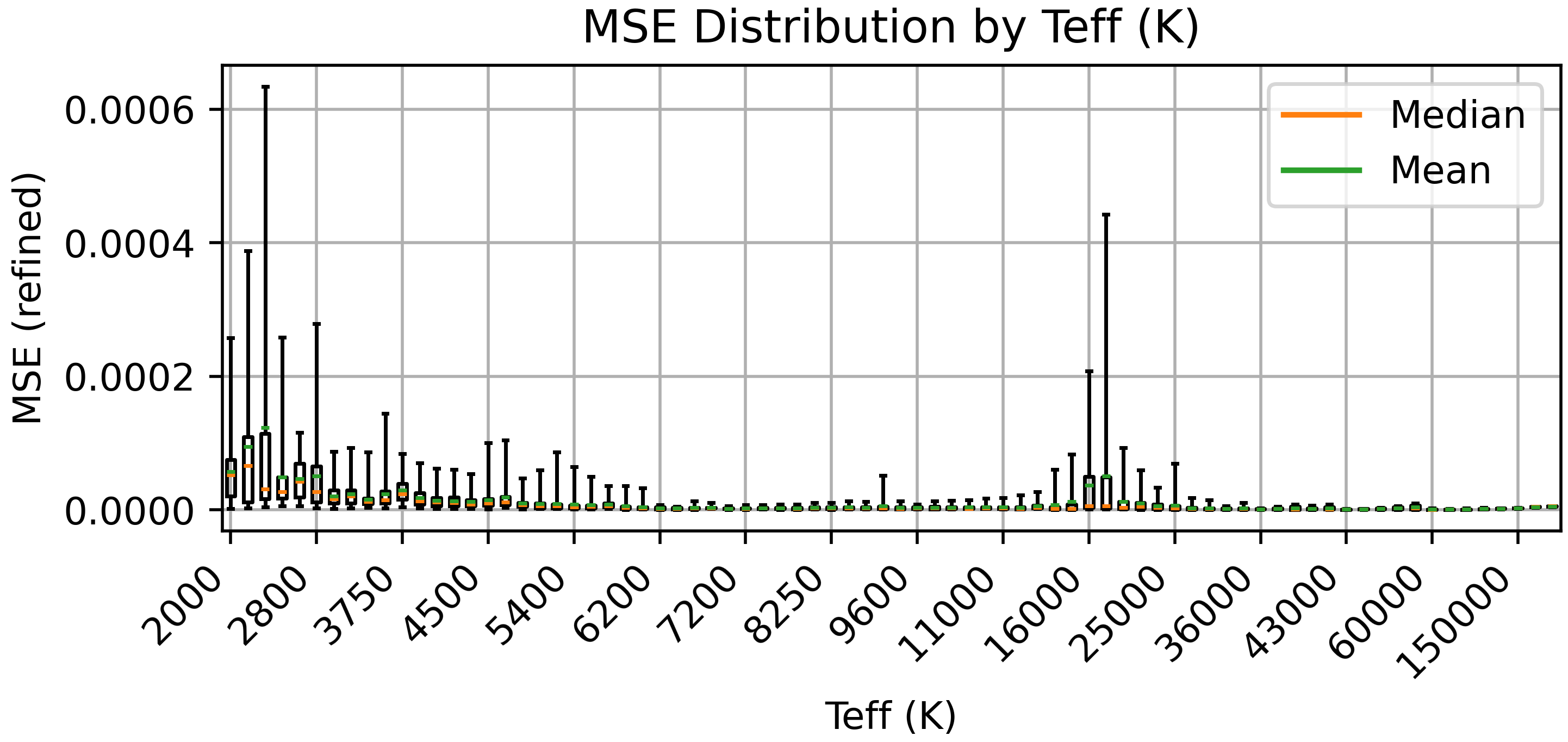}
        \caption{Training set}
    \end{subfigure}
    \begin{subfigure}{1\textwidth}
        \centering
        \includegraphics[width=\textwidth]{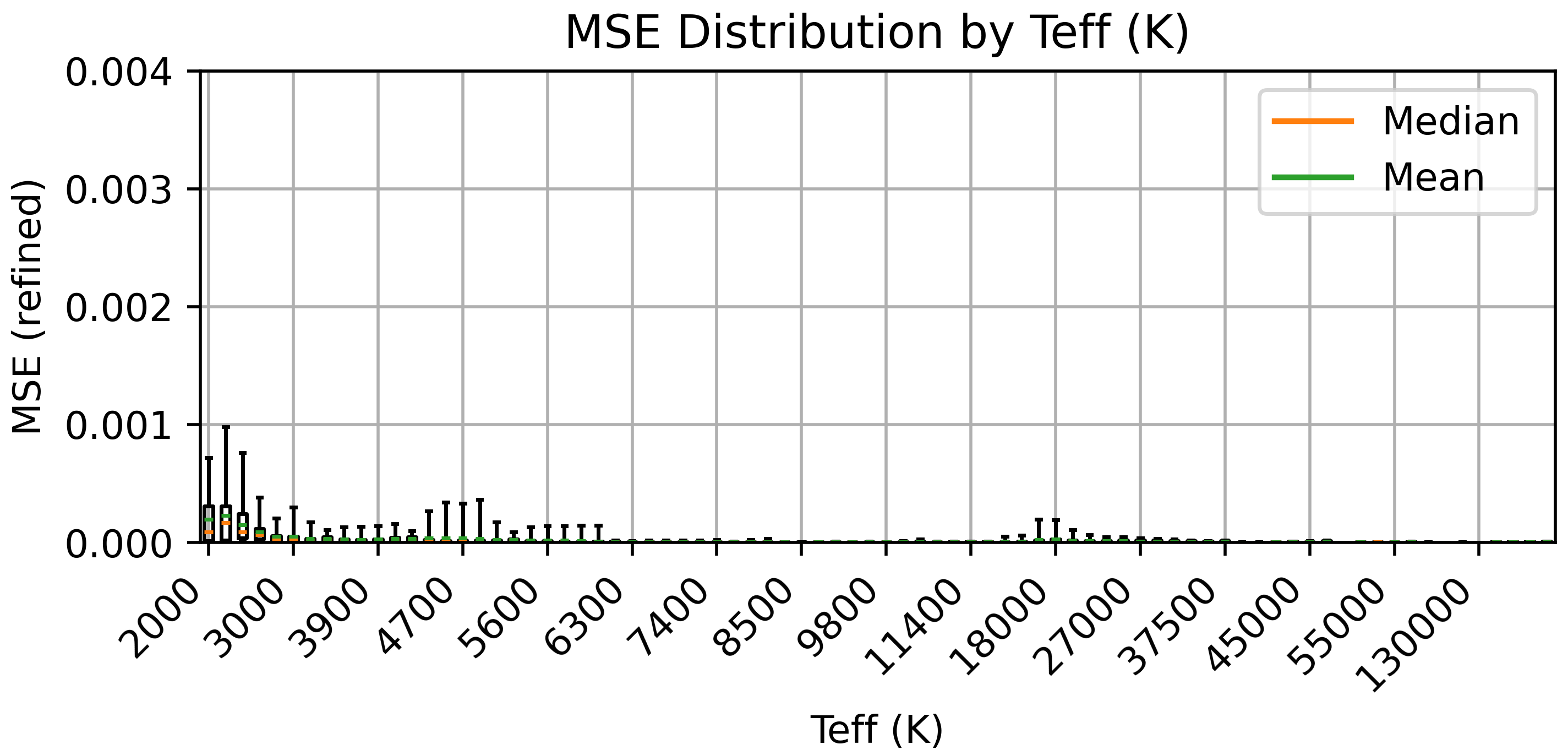}
        \caption{Validation set}
    \end{subfigure}
    \begin{subfigure}{1\textwidth}
        \centering
        \includegraphics[width=\textwidth]{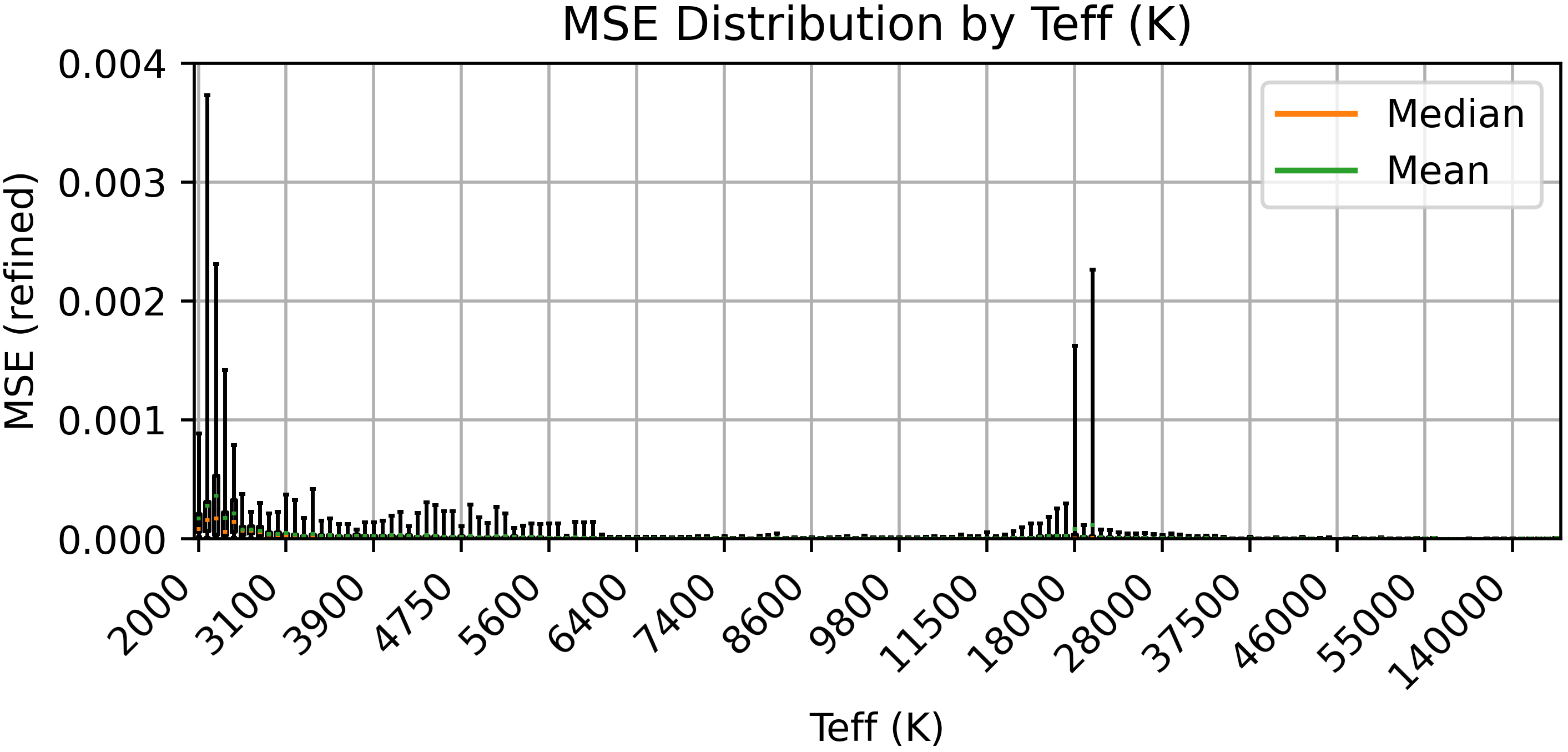}
        \caption{Test set}
    \end{subfigure}
    \caption{%
        Distribution of MSE prediction errors across $T_{\mathrm{eff}}$.  
        High MSE occurs predominantly at low temperatures, with consistent trends across splits, indicating robust generalisation.}
    \label{fig:box_teff}
\end{figure}

\begin{figure}[!b]
    \centering

    \begin{subfigure}{1\textwidth}
        \centering
        \includegraphics[width=\textwidth]{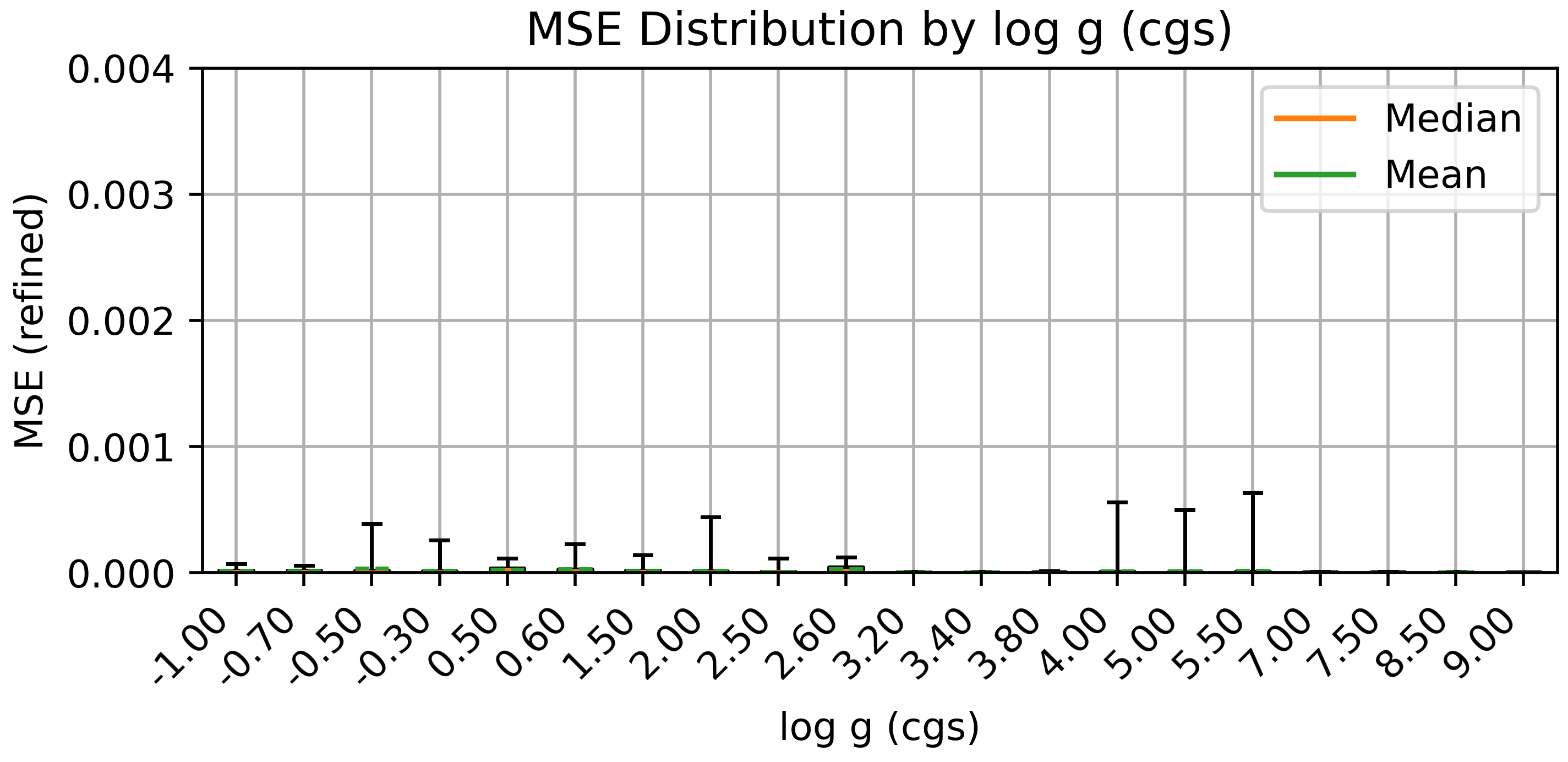}
        \caption{Training set}
    \end{subfigure}
    \begin{subfigure}{1\textwidth}
        \centering
        \includegraphics[width=\textwidth]{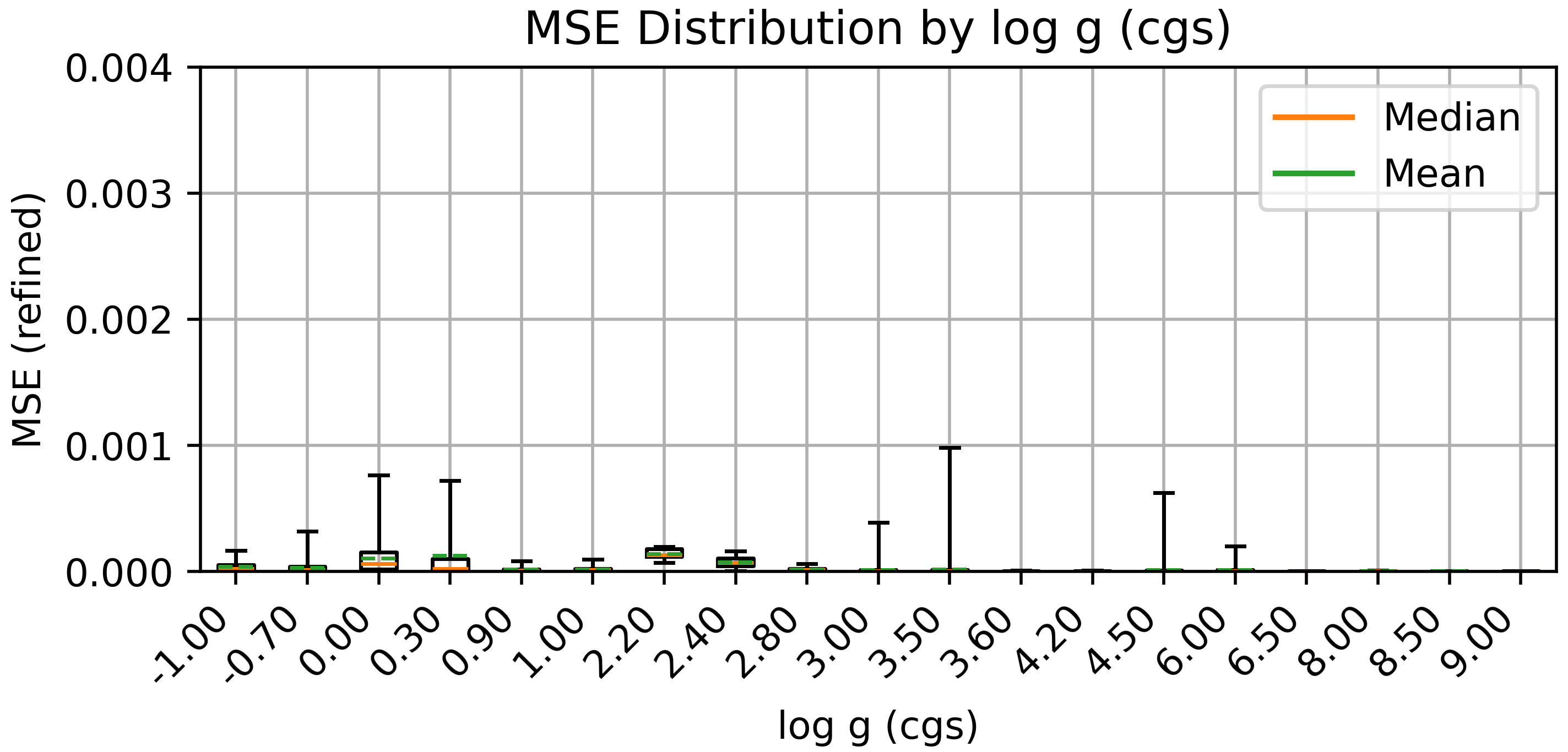}
        \caption{Validation set}
    \end{subfigure}
    \begin{subfigure}{1\textwidth}
        \centering
        \includegraphics[width=\textwidth]{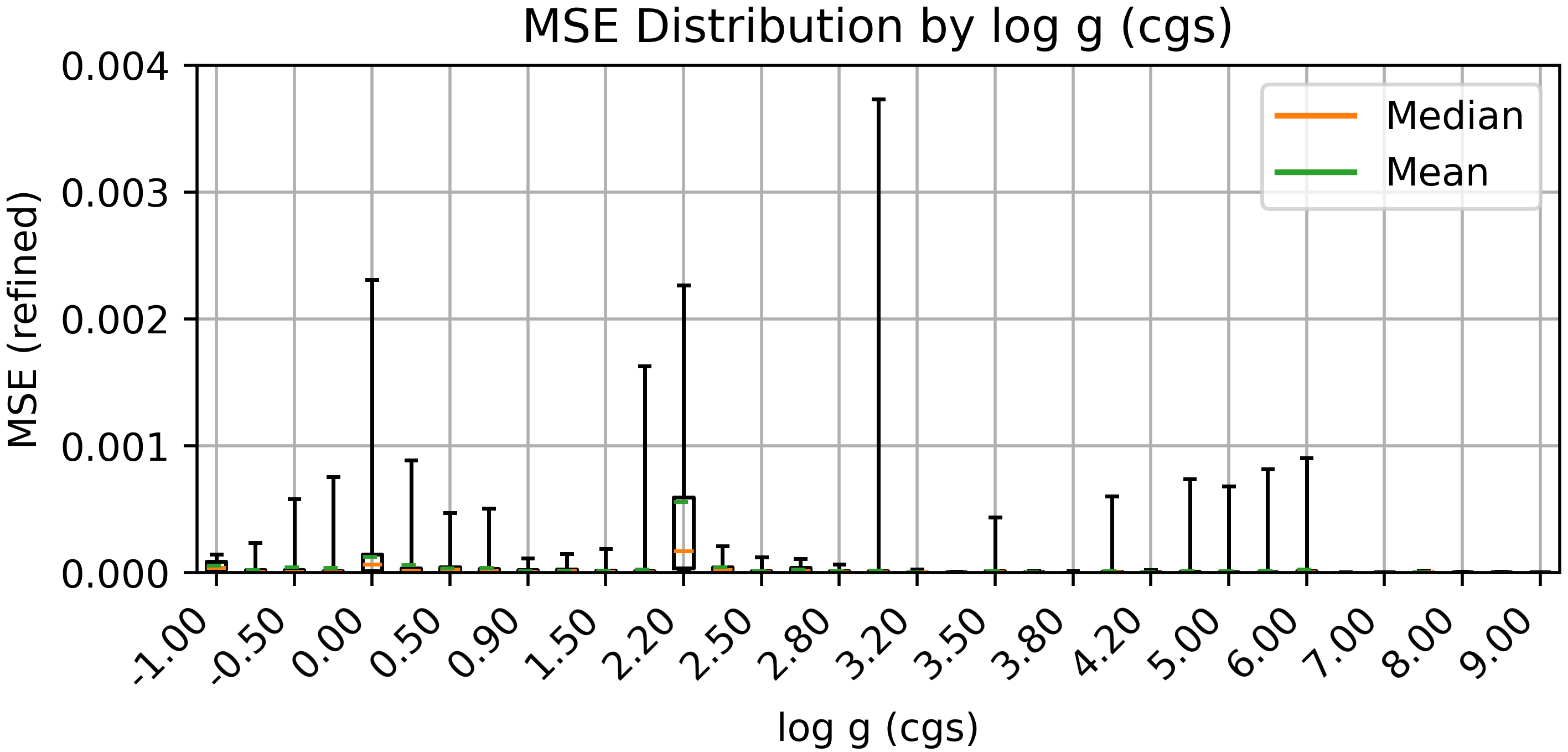}
        \caption{Test set}
    \end{subfigure}

    \caption{%
        Distribution of MSE across $\log g$.  
        Elevated errors appear near the boundaries of PHOENIX--Husser and C3K overlap, indicating inconsistencies between the underlying physics of the two libraries.}
    
    \label{fig:box_logg}
\end{figure}

\begin{figure}[!t]
    \centering

    \begin{subfigure}{1\textwidth}
        \centering
        \includegraphics[width=\textwidth]{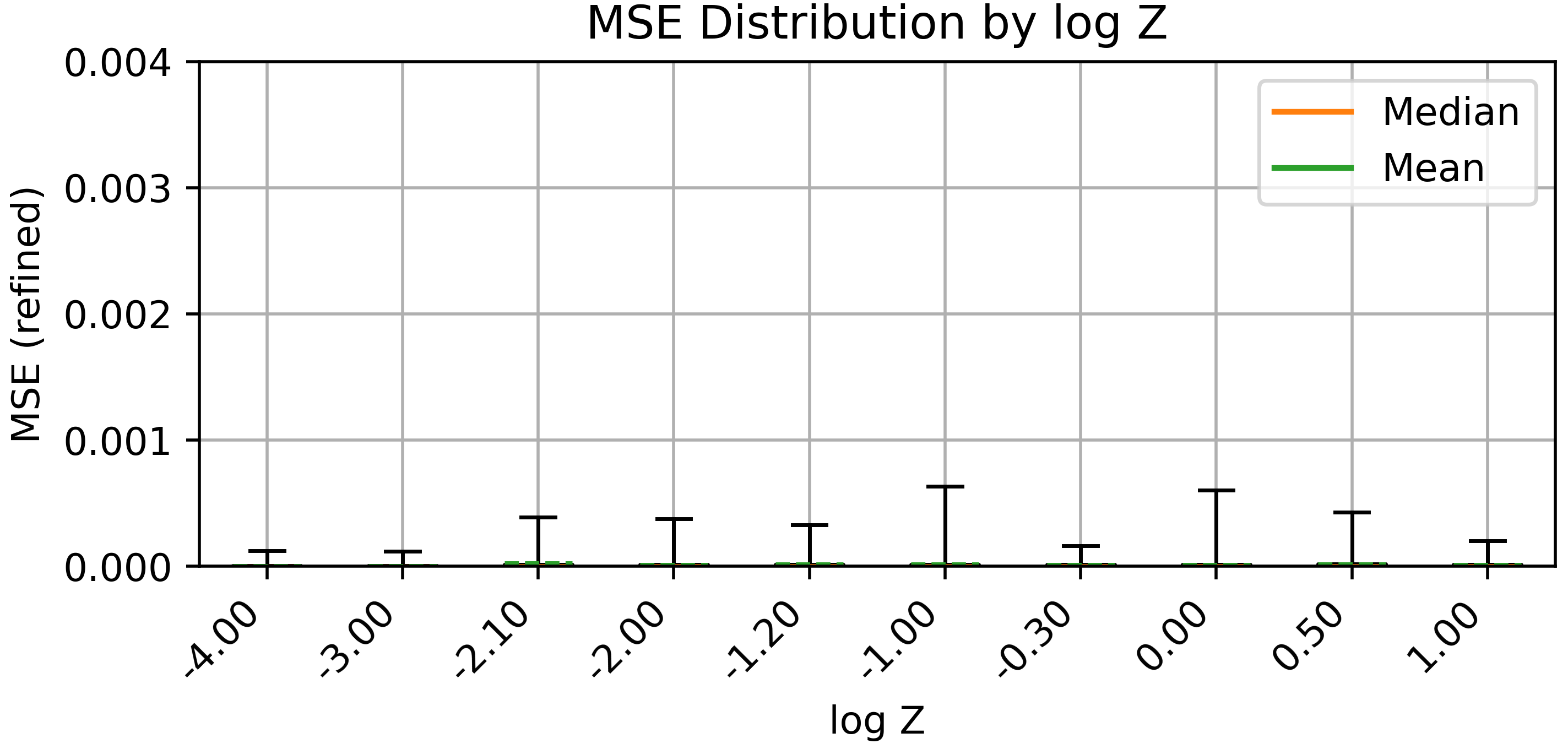}
        \caption{Training set}
    \end{subfigure}
    \begin{subfigure}{1\textwidth}
        \centering
        \includegraphics[width=\textwidth]{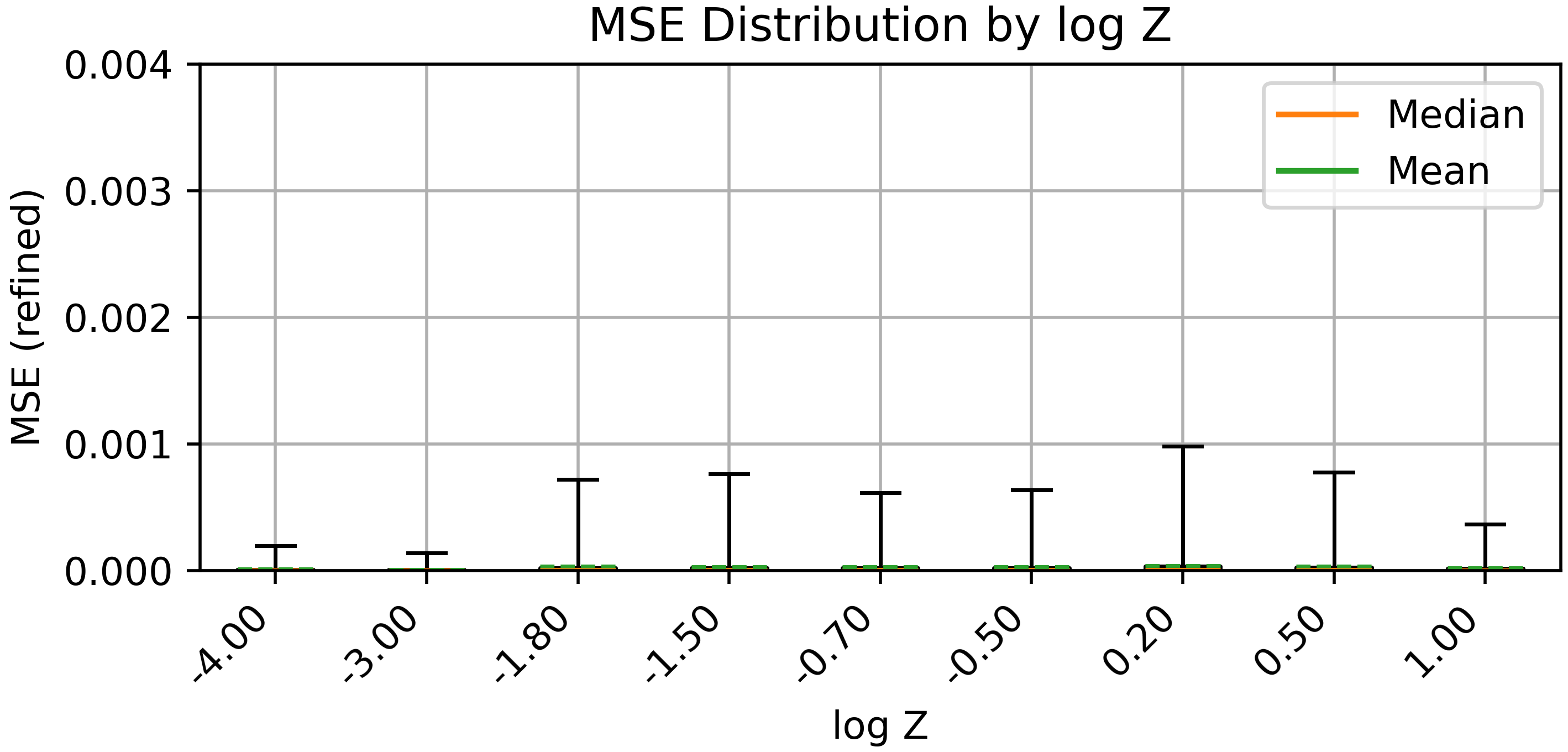}
        \caption{Validation set}
    \end{subfigure}
    \begin{subfigure}{1\textwidth}
        \centering
        \includegraphics[width=\textwidth]{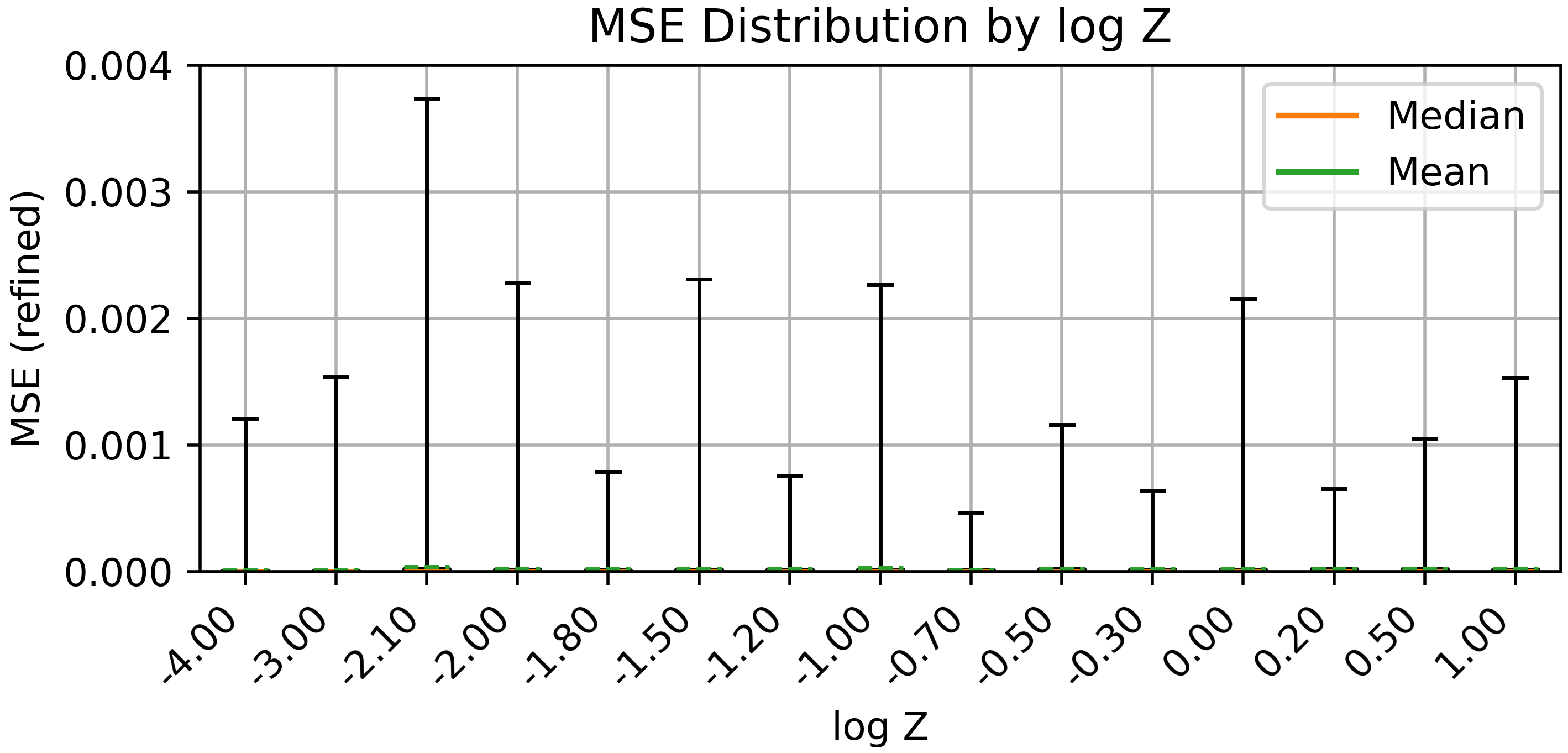}
        \caption{Test set}
    \end{subfigure}

    \caption{%
        Distribution of MSE across $\log Z$.  
        Errors are broadly uniform, with particularly low mean and median MSEs at $\log Z=-4$ and $-3$, which originate from regions governed by a single library and therefore contain no cross-library inconsistencies.}
    \label{fig:box_logZ}
\end{figure}

\begin{figure}[!t]
    \centering

    \begin{subfigure}{1\textwidth}
        \centering
        \includegraphics[width=\textwidth]{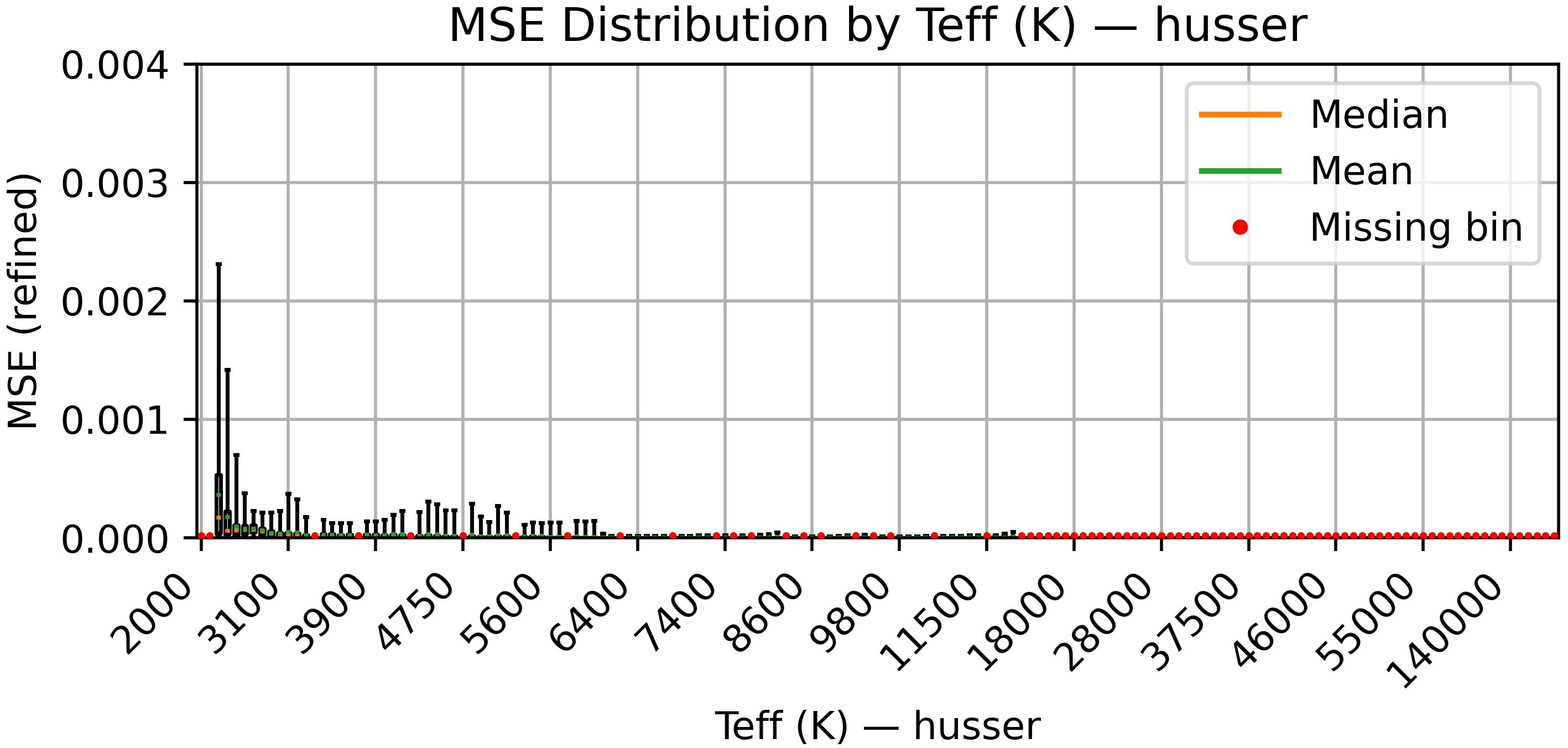}
    \end{subfigure}
    \begin{subfigure}{1\textwidth}
        \centering
        \includegraphics[width=\textwidth]{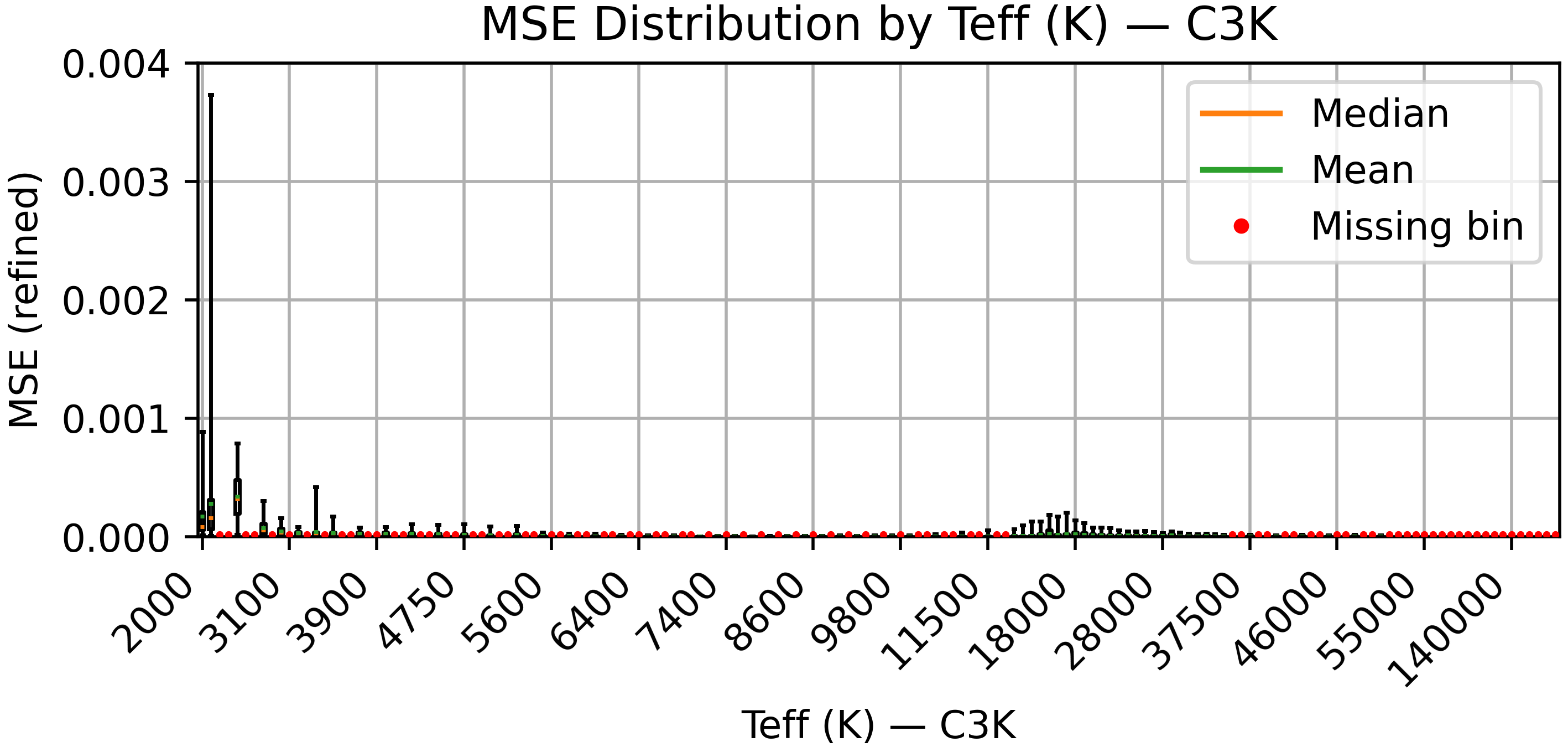}
    \end{subfigure}
    \begin{subfigure}{1\textwidth}
        \centering
        \includegraphics[width=\textwidth]{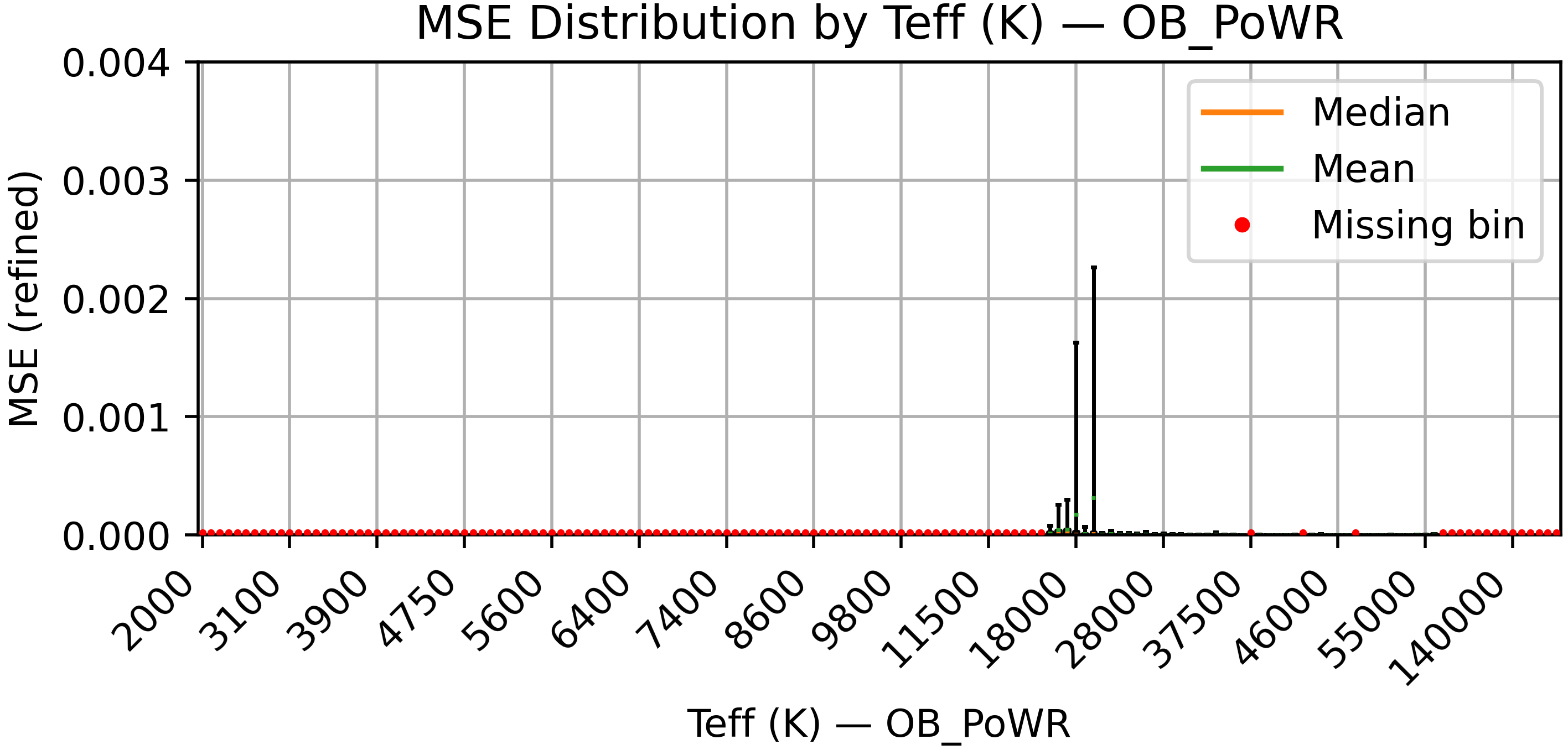}
    \end{subfigure}
    \begin{subfigure}{1\textwidth}
        \centering
        \includegraphics[width=\textwidth]{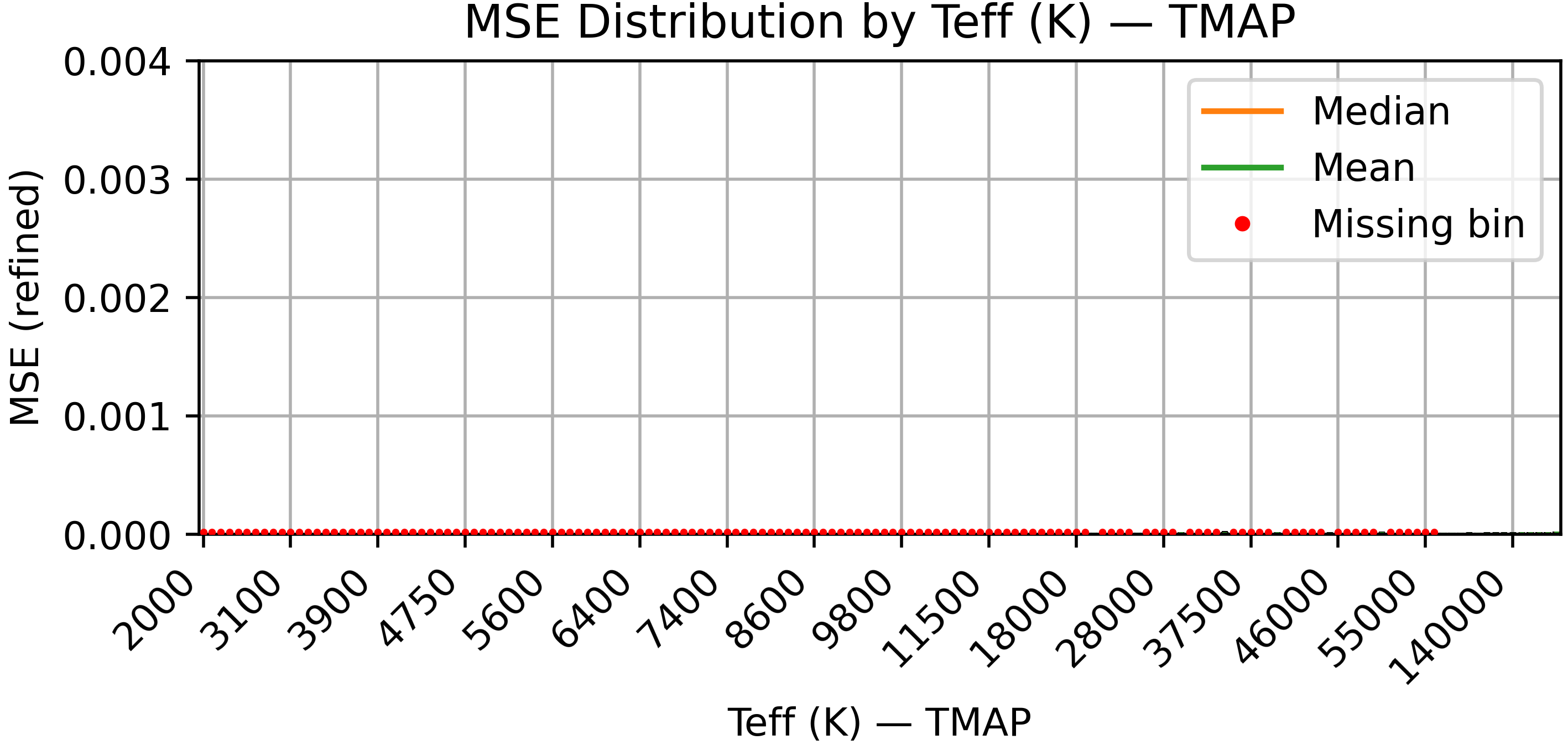}
    \end{subfigure}

    \caption{%
        Distribution of MSE across $T_{\mathrm{eff}}$ for the four libraries.}
    \label{fig:box_teff_test_lib}
\end{figure}

\begin{figure}[!t]
    \centering

    \begin{subfigure}{1\textwidth}
        \centering
        \includegraphics[width=\textwidth]{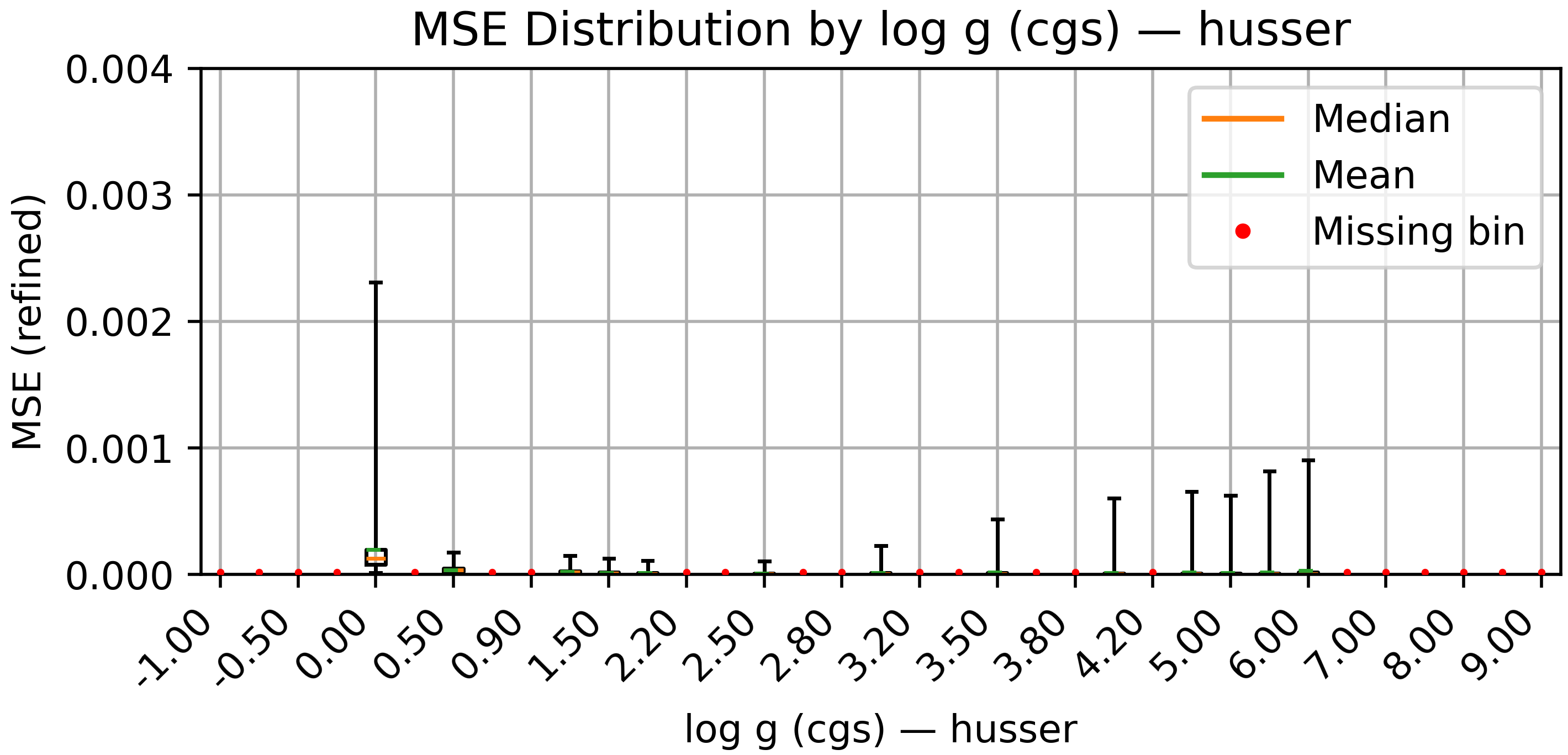}
    \end{subfigure}
    \begin{subfigure}{1\textwidth}
        \centering
        \includegraphics[width=\textwidth]{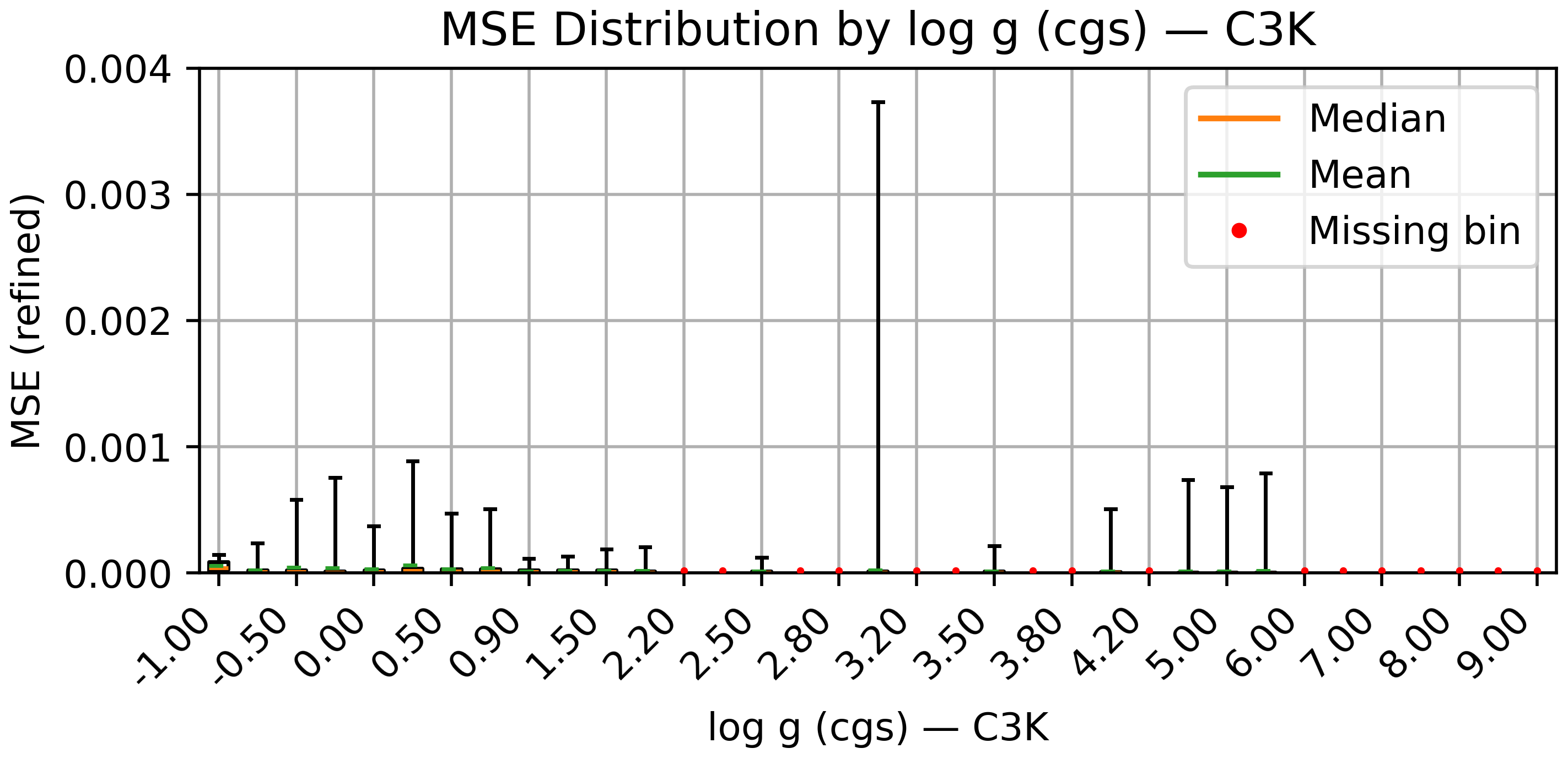}
    \end{subfigure}
    \begin{subfigure}{1\textwidth}
        \centering
        \includegraphics[width=\textwidth]{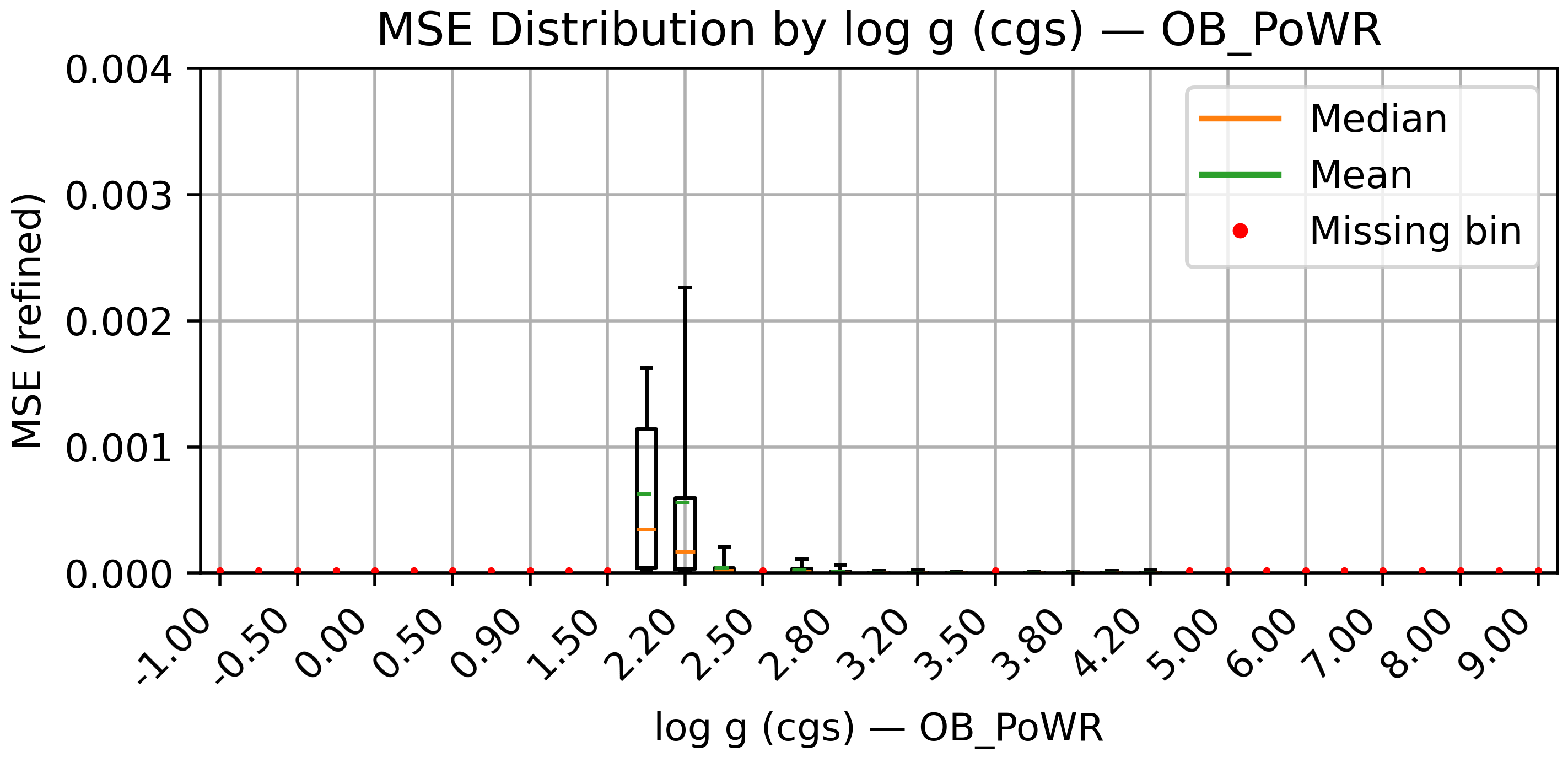}
    \end{subfigure}
    \begin{subfigure}{1\textwidth}
        \centering
        \includegraphics[width=\textwidth]{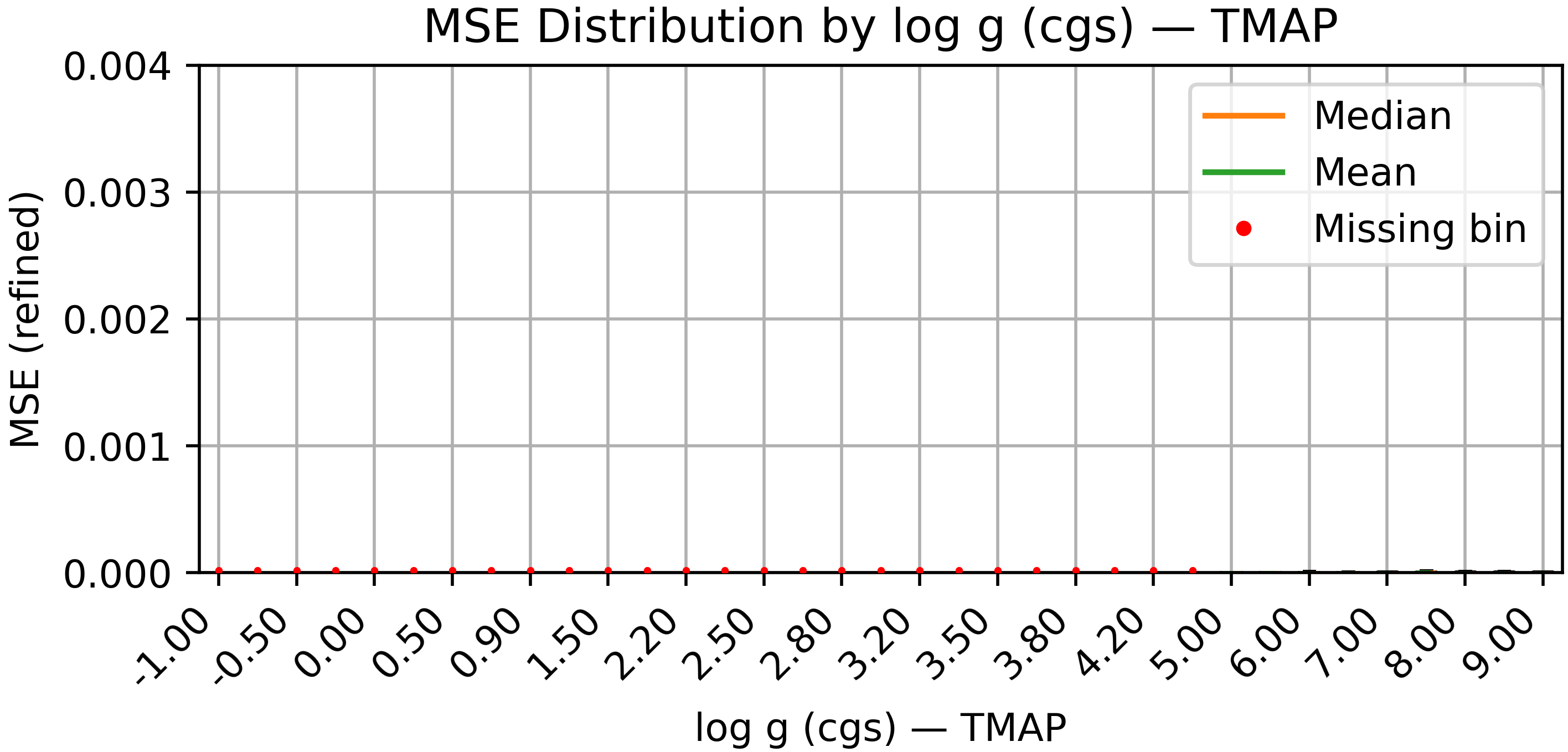}
    \end{subfigure}

    \caption{%
        Distribution of MSE across $\log g$ for the four libraries.}
    \label{fig:box_logg_test_lib}
\end{figure}

\begin{figure}[!t]
    \centering
    \begin{subfigure}{1\textwidth}
        \centering
        \includegraphics[width=\textwidth]{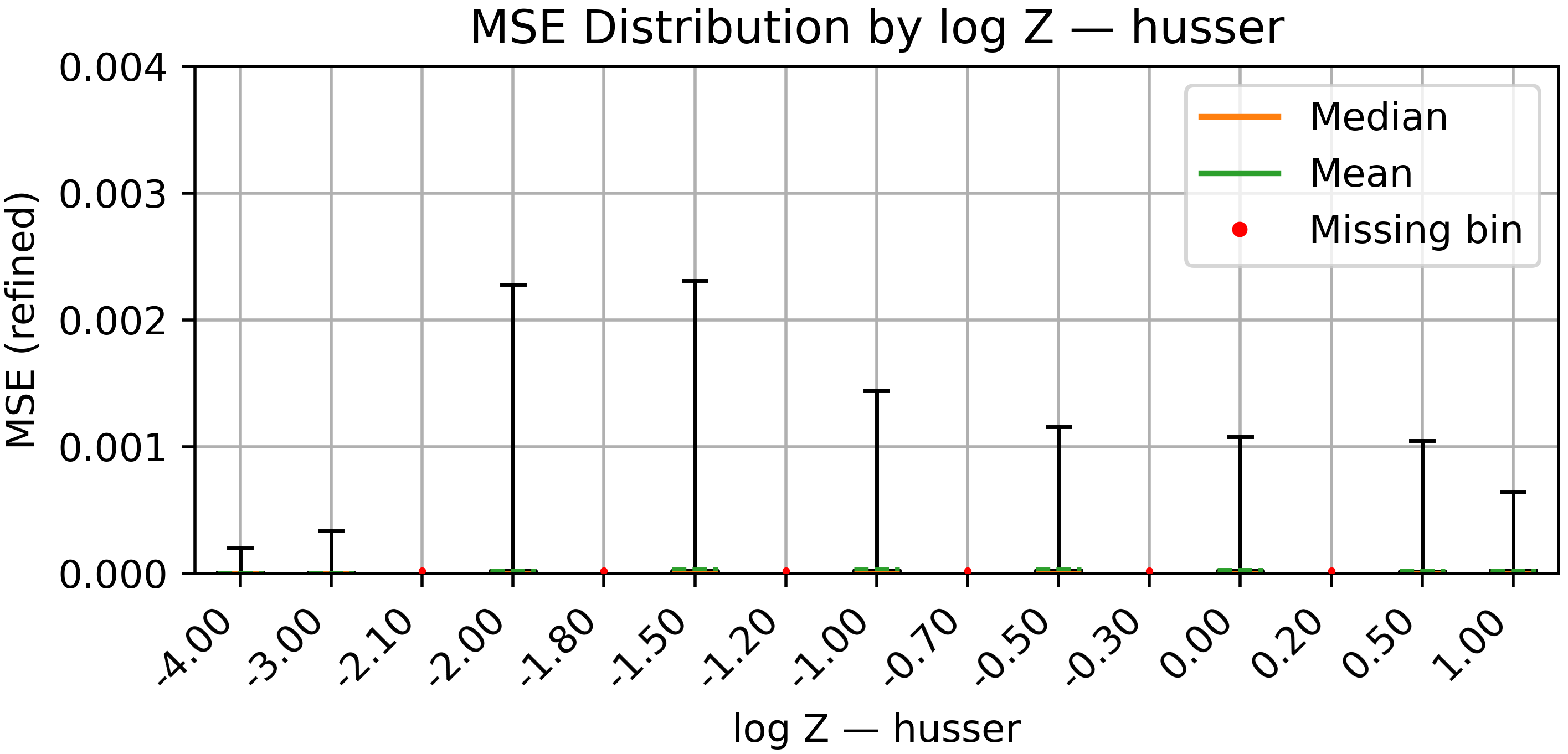}
    \end{subfigure}
    \begin{subfigure}{1\textwidth}
        \centering
        \includegraphics[width=\textwidth]{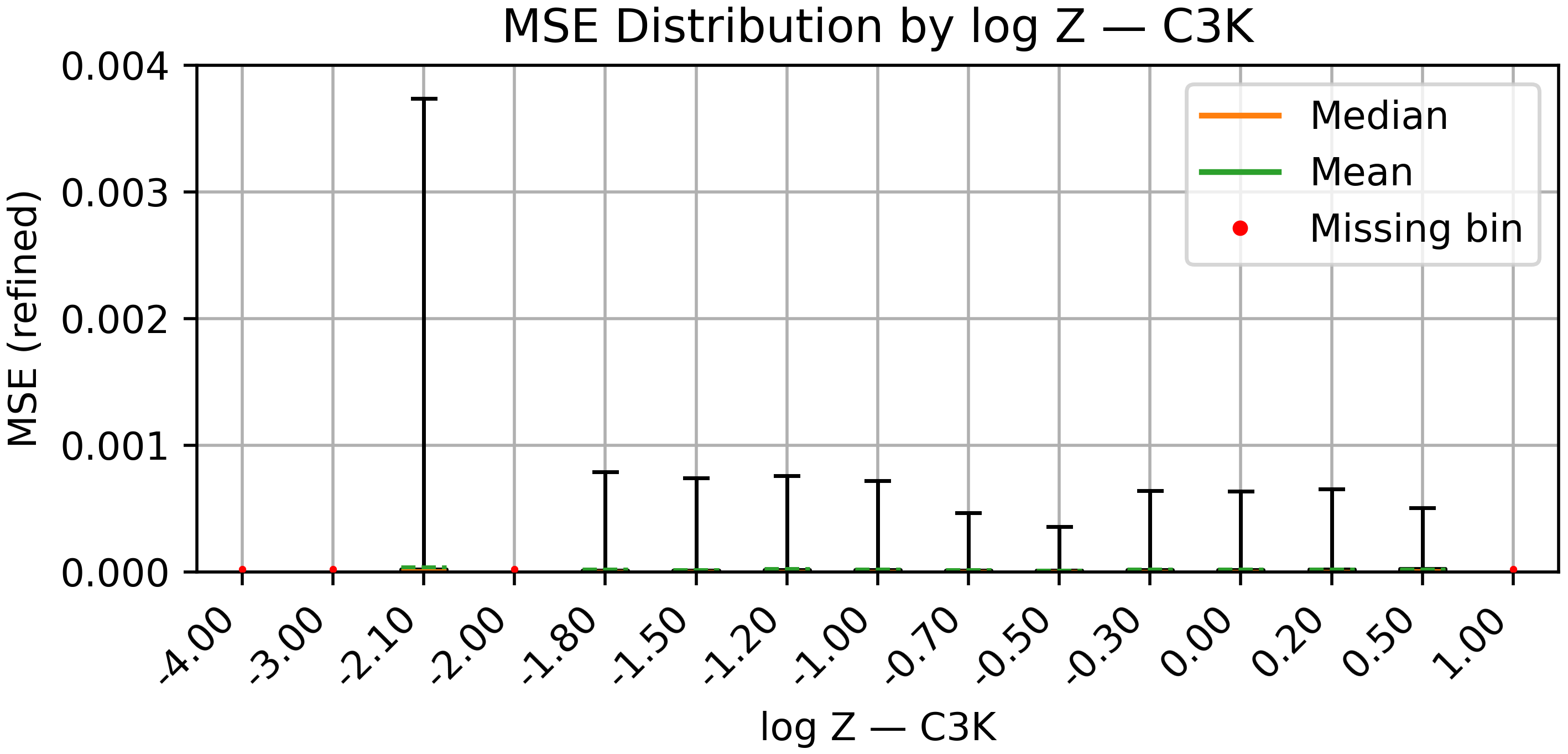}
    \end{subfigure}
    \begin{subfigure}{1\textwidth}
        \centering
        \includegraphics[width=\textwidth]{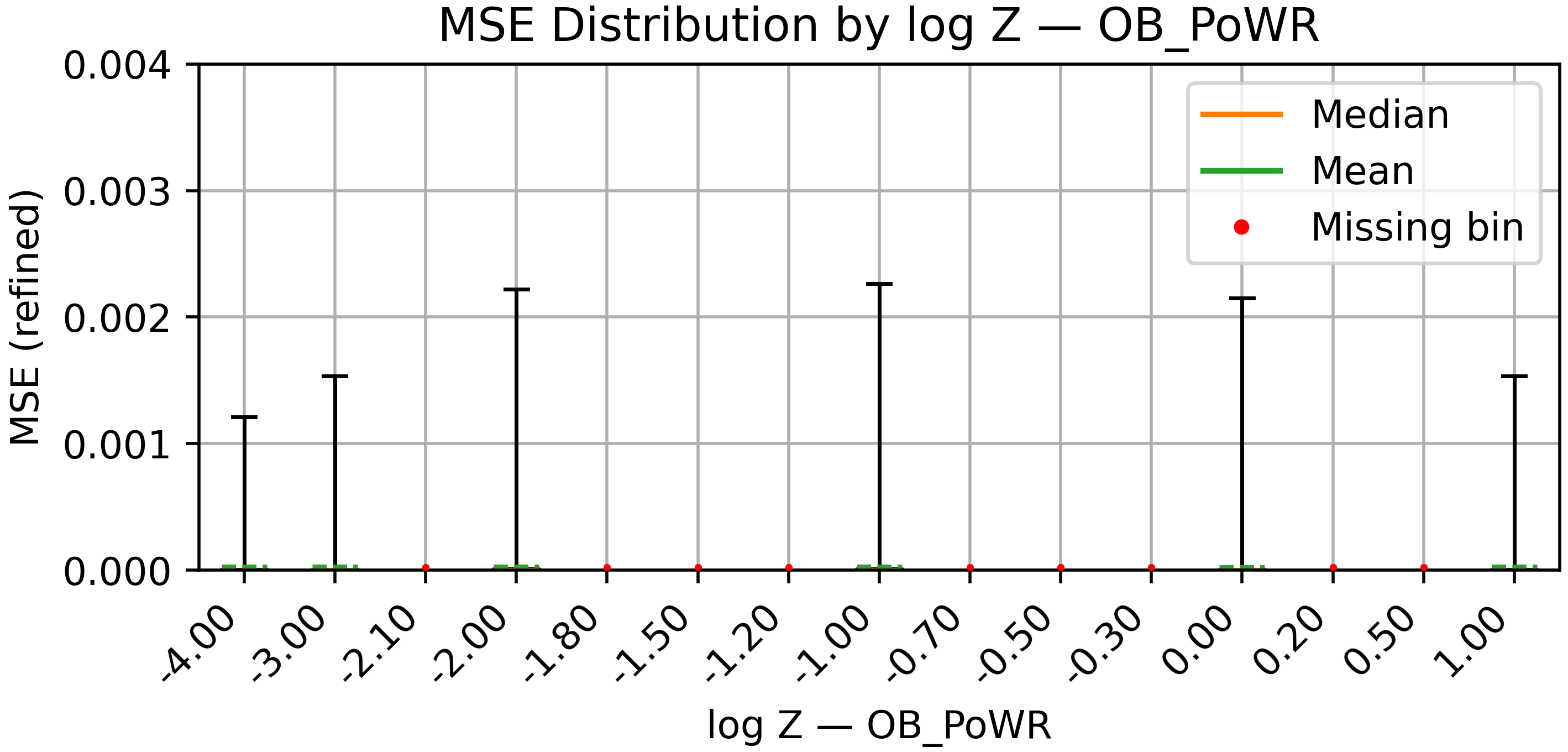}
    \end{subfigure}
    \begin{subfigure}{1\textwidth}
        \centering
        \includegraphics[width=\textwidth]{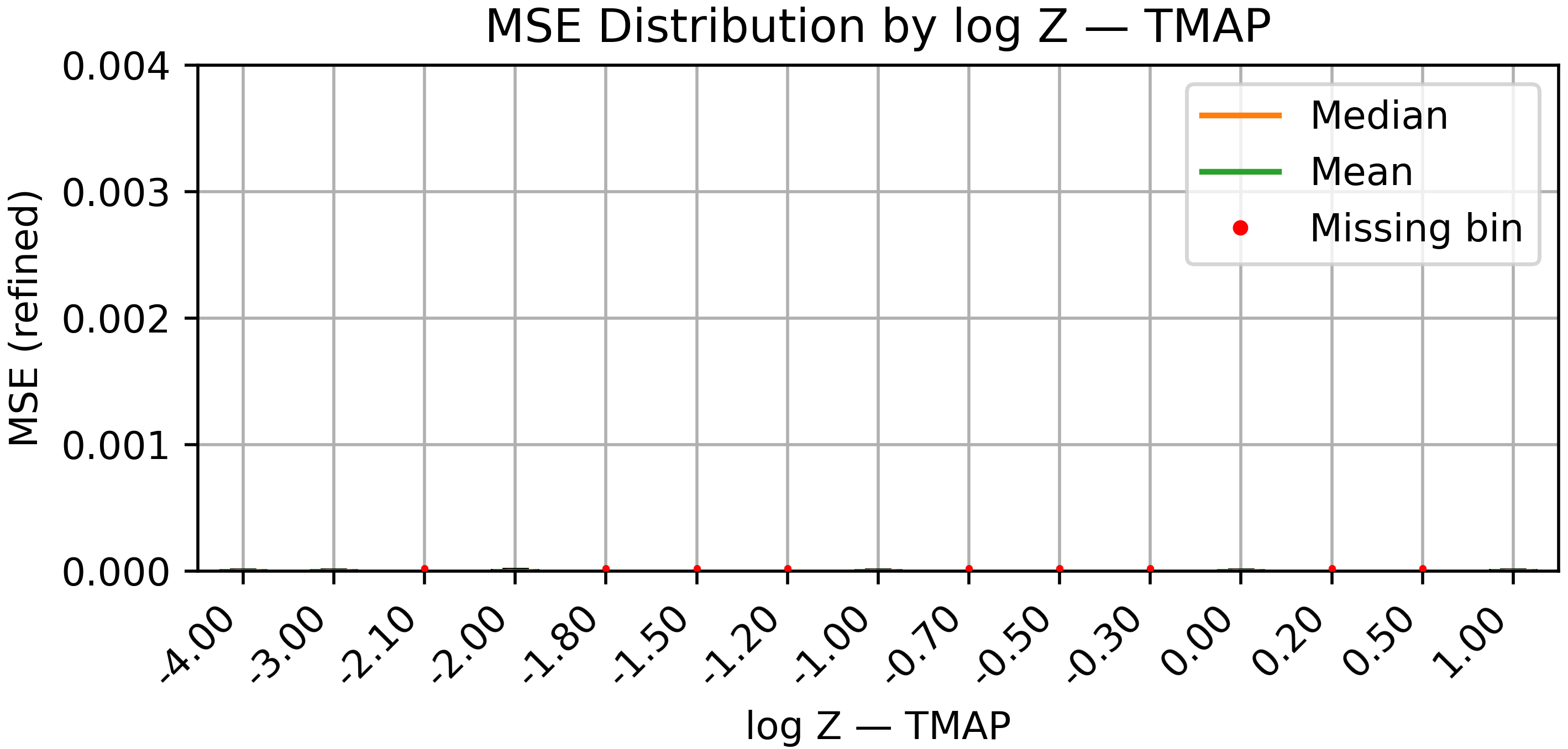}
    \end{subfigure}

    \caption{%
        Distribution of MSE across $\log Z$ for the four libraries.}
    \label{fig:box_logz_test_lib}
\end{figure}

In this section, we evaluate the quantitative and qualitative reconstruction performance of SM-Net across the training, validation, and test sets.
All spectra are represented in the log1p-scaled flux space as described in Section~2, and this transformation is not inverted at any stage of training or evaluation.
Consequently, all residuals and mean squared error (MSE) values reported here are computed directly in the log1p-scaled space and should not be interpreted as errors in linear flux units.

First, we analyse the reconstruction accuracy as a function of the three physical parameters $(T_{\mathrm{eff}}, \log g, \log Z)$ across the training, validation, and test splits.
As discussed in Section~2, spectra from the TMAP and OB-PoWR libraries were replicated across discrete $\log Z$ values to enable embedding within the joint parameter space.
In metallicity bins populated exclusively by these libraries, the spectra are therefore invariant by construction with respect to $\log Z$ and do not probe metallicity-dependent spectral sensitivity.
This limitation should be borne in mind when interpreting the error distributions shown in Figures~\ref{fig:box_teff}, \ref{fig:box_logg}, and \ref{fig:box_logZ}.


Table~\ref{tab:smnet_quality} summarises performance across the three dataset splits. The mean per-spectrum MSE values are $1.47\times10^{-5}$ (train), $2.62\times10^{-5}$ (validation), and $2.34\times10^{-5}$ (test), corresponding to root-mean-square deviations of $0.38$--$0.51$ per cent in the transformed log1p-scaled flux space used for training and evaluation. The median MSE values of $(4.2$--$5.6)\times10^{-6}$ further indicate that the \emph{majority} of spectra are reconstructed with better than $0.5\%$ RMS accuracy in that transformed representation.

The 10th-percentile MSE values lie in the range $(1.1$–$1.6)\times10^{-6}$, corresponding to RMS deviations of $\sim0.1\%$.
These spectra represent the best-reconstructed 10\% of cases, in which SM-Net achieves extremely tight agreement with the ground truth.
The 90th-percentile values, $(3.17$–$5.35)\times10^{-5}$, correspond to RMS errors of $\sim1\%$.
This demonstrates that even the upper tail of the error distribution remains modest and that SM-Net generalises reliably across non-uniform regions of the combined grid.

The minimum MSE values reach as low as $(1.8$–$2.6)\times10^{-7}$, equivalent to RMS flux errors below $0.05\%$.
These near-perfect reconstructions occur primarily for spectra with $T_{\mathrm{eff}} \geq 40{,}000$~$\mathrm{K}$, $\log g \geq 6$, and $\log Z \leq -3$
(Figures~\ref{fig:box_teff}--\ref{fig:box_logZ}), where the grid coverage originates from a single library and contains no cross-library physics mismatches.

Conversely, the maximum MSE values are in the range of $(6.3$–$9.8)\times10^{-4}$ for train/validation and $3.73\times10^{-3}$ for test, corresponding to 2--6\% RMS errors.
These are genuine worst-case outliers and occur almost exclusively in the lowest-temperature region of the combined grid, particularly below $T_{\mathrm{eff}}\approx3,000$~$\mathrm{K}$.
These isolated high-error cases reflect intrinsic difficulty rather than systematic model failure.

The validation split is particularly important and challenging, as, in addition to boundary overlaps, \emph{none} of the three input parameters in this split appear in training.
The close agreement between validation and test errors, therefore, demonstrates that SM-Net successfully interpolates across unseen regions of the stellar parameter grid, a core objective of this work.

\subsection{Error Distributions Across the Parameter Space}

To diagnose potential biases on the non-uniform combined grid, Figures~\ref{fig:box_teff}, \ref{fig:box_logg}, and \ref{fig:box_logZ} show the MSE distribution as a function of $T_{\mathrm{eff}}$, $\log g$, and $\log Z$ parameters across all splits, respectively. One can draw the following conclusions based on these MSE box plots.



The train, validation, and test splits exhibit very similar MSE profiles as functions of Teff, log g, and log Z. Although the absolute values differ slightly, the overall shapes of the distributions and the locations of the outliers are consistent across all three splits. Together with the close agreement in the summary statistics in Table~\ref{tab:smnet_quality}, this indicates that SM-Net does not overfit and generalises robustly across unseen regions of the parameter grid.

The largest errors and most prominent outliers occur at the lowest temperatures (roughly 2,000~$\mathrm{K}$ $\leq $\,$T_{\mathrm{eff}} \leq 3,000$\,~$\mathrm{K}$), where spectra are both more complex and less uniformly behaved across the source libraries. Elevated outlier rates are also seen near parameter ranges where libraries overlap, especially around the PHOENIX–Husser / C3K transition, consistent with the cross-library mismatches discussed in Section 5. By contrast, regions dominated by a single internally consistent library, such as the hottest TMAP regime or the most metal-poor PHOENIX–Husser bins, show uniformly low errors and minimal scatter.



The box plots in Figures~\ref{fig:box_teff_test_lib}, \ref{fig:box_logg_test_lib}, and \ref{fig:box_logz_test_lib} show the distribution of MSE values for the test set, stratified by stellar parameter and separated by the four constituent libraries in the combined dataset.
Bins for which a given library provides no spectra are indicated by a red marker.
For TMAP, the reconstruction errors are systematically much smaller than those of the other libraries.
Because a common vertical scale is used across all panels to enable direct comparison, the TMAP error distributions are compressed near zero, which may make them appear visually sparse or absent.
A closer inspection shows that this reflects genuinely negligible reconstruction errors rather than missing data or failed reconstructions.

\begin{figure}[!b]
    \centering
    \includegraphics[width=1\textwidth]{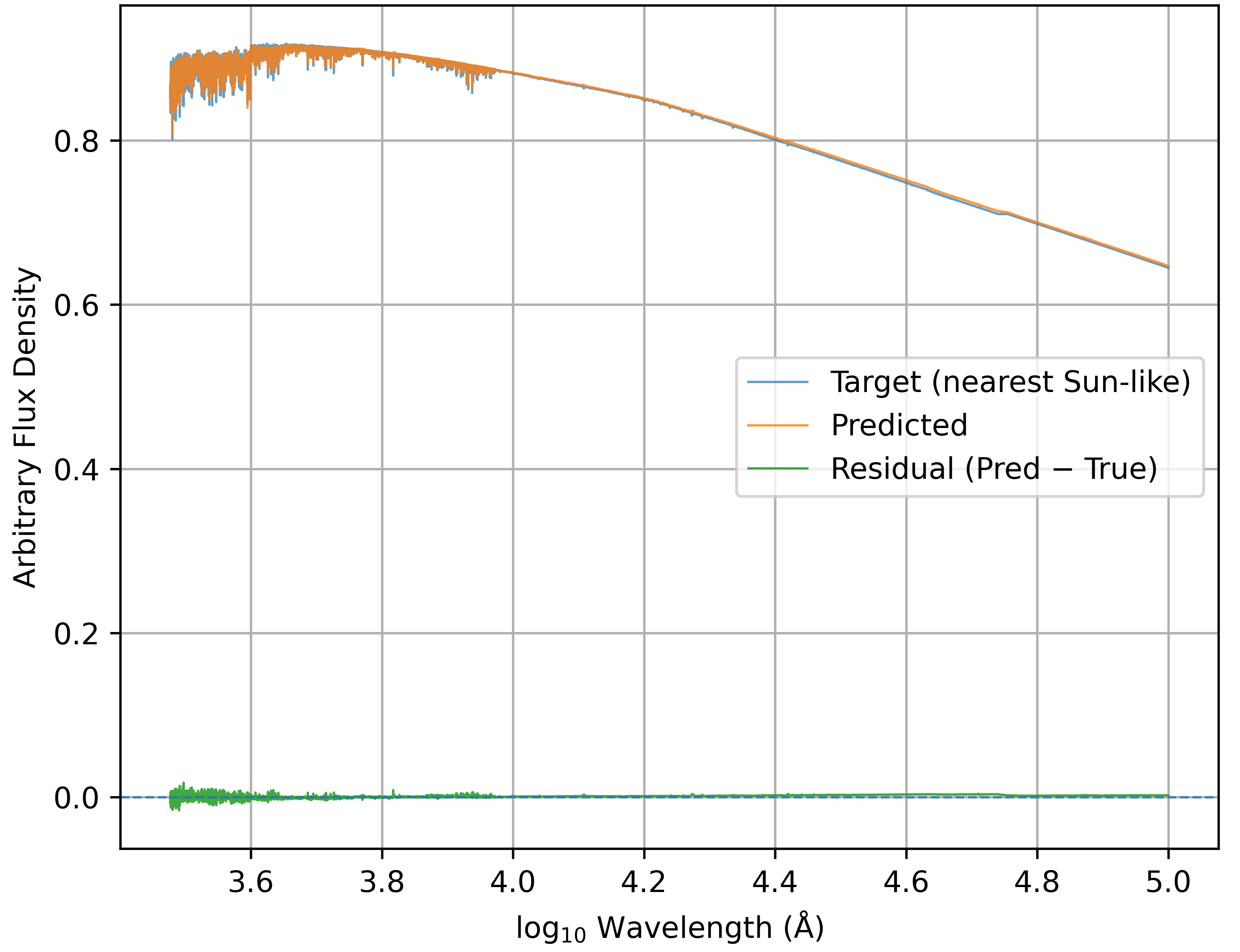}
    \caption{
    Predicted spectrum for Sun, and the target spectrum from PHOENIX-Husser with the closest Sun-like parameters.}
    \label{fig:Sun}
\end{figure}

\begin{figure}[!t]
    \centering
    \includegraphics[width=1\textwidth]{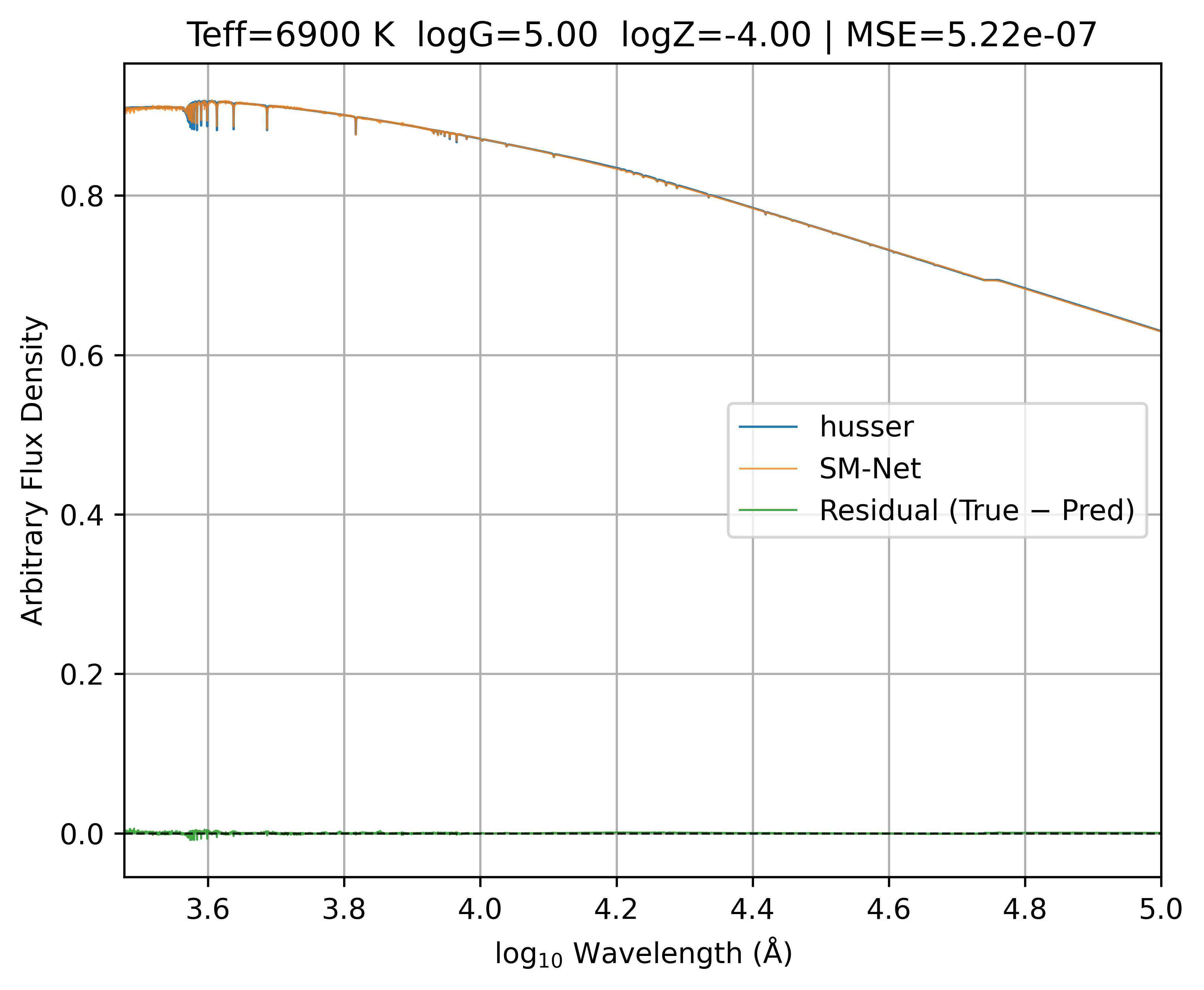}
    \caption{
    High-fidelity SM-Net reconstruction for
    $T_{\mathrm{eff}} = 6900,\mathrm{K}$, $\log g = 5.0$, and $\log Z = -4.0$,
    shown against the corresponding PHOENIX–Husser reference spectrum.}
    \label{fig:good1}
\end{figure}

\begin{figure}[!b]
    \centering
    \includegraphics[width=1\textwidth]{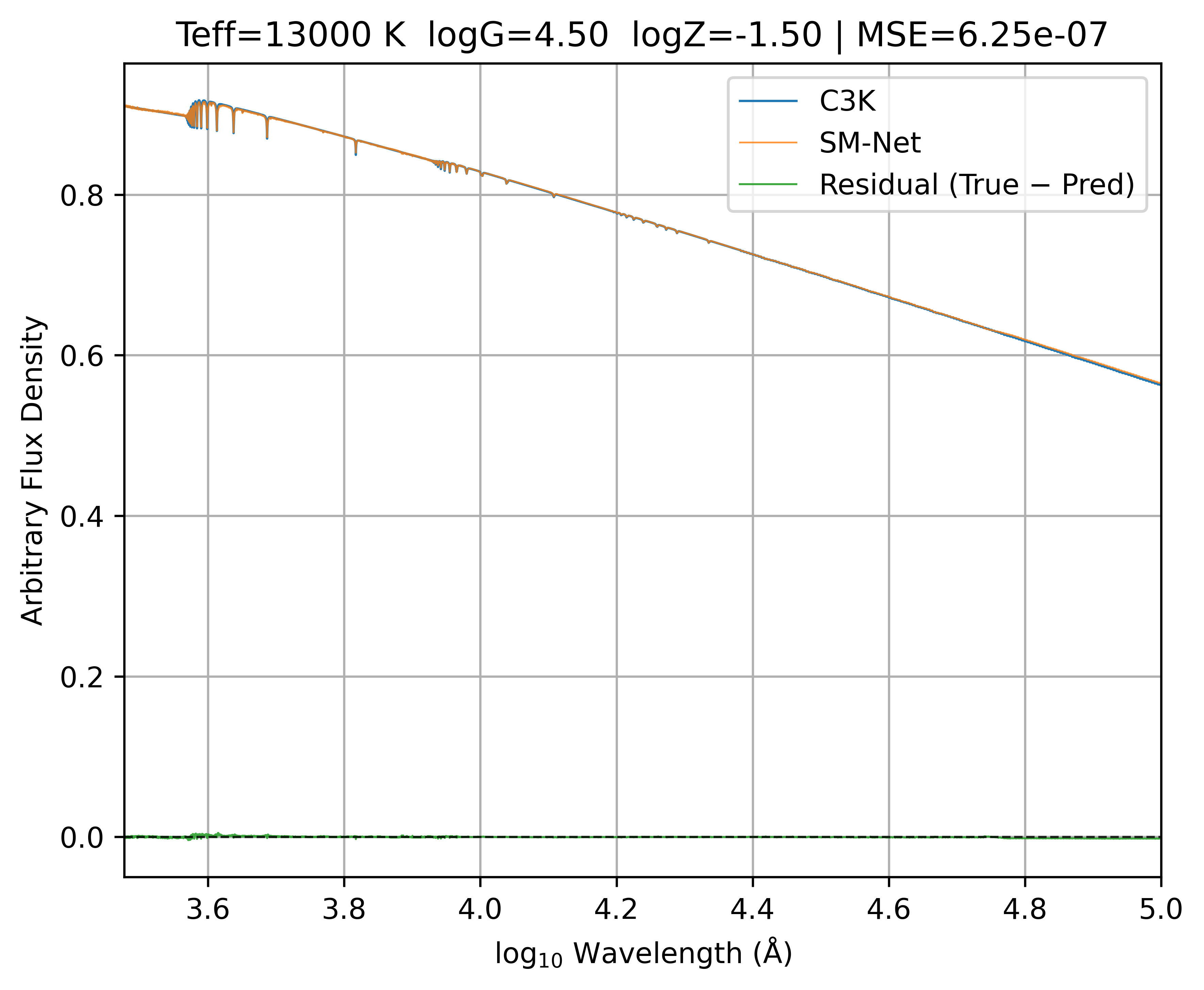}
    \caption{
    High-quality SM-Net reconstruction for
    $T_{\mathrm{eff}} = 13{,}000~\mathrm{K}$, $\log g = 4.5$, and $\log Z = -1.5$,
    compared with the corresponding C3K reference spectrum.}
    \label{fig:good2}
\end{figure}

\begin{figure}[!t]
    \centering
    \includegraphics[width=1\textwidth]{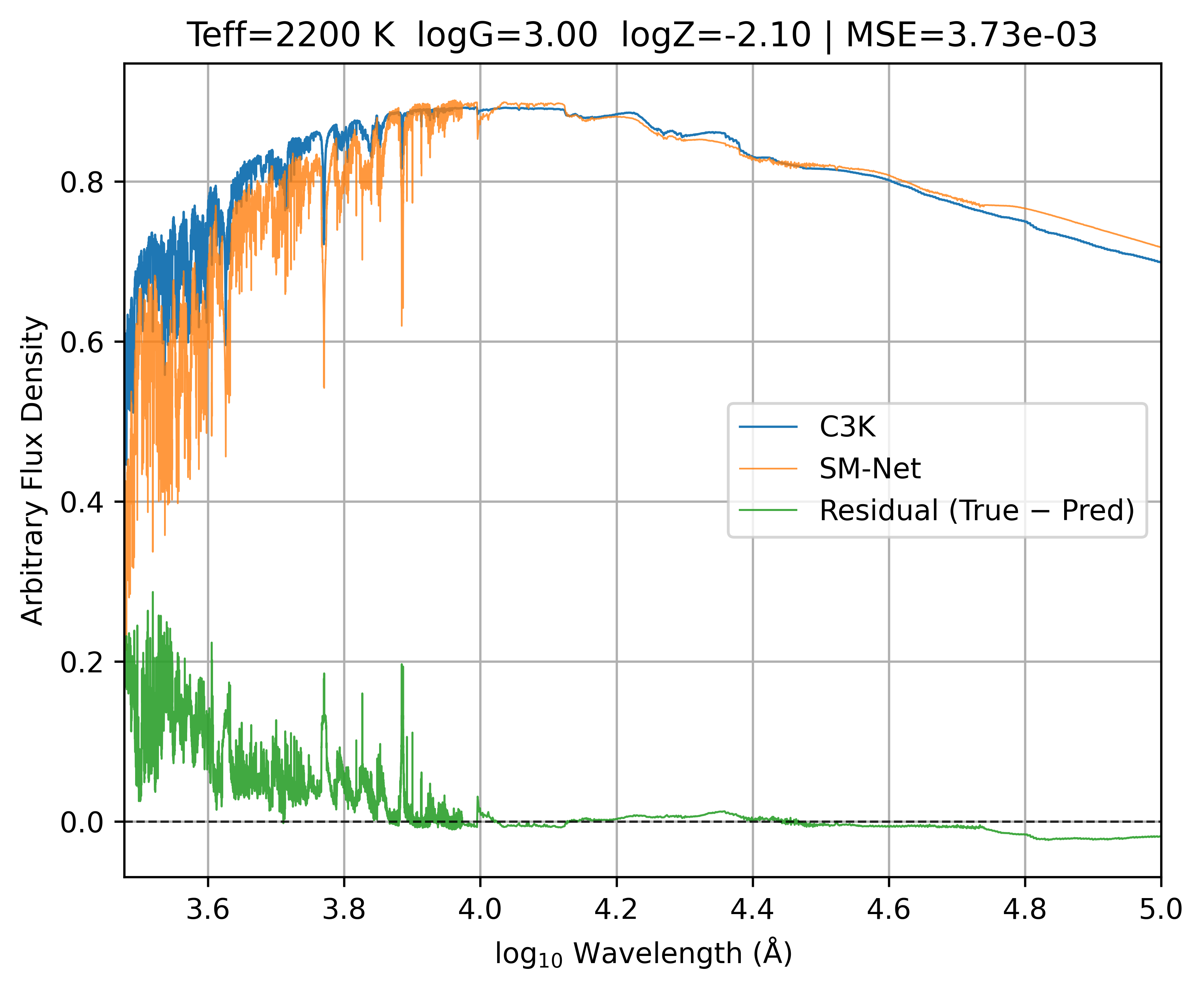}
    \caption{
    Example of a low-quality reconstruction at
    $T_{\mathrm{eff}} = 2,200~\mathrm{K}$, $\log g = 3.0$, and $\log Z = -2.1$,
    shown against the corresponding C3K reference spectrum.}
    \label{fig:bad1}
\end{figure}

\begin{figure}[!b]
    \centering
    \includegraphics[width=1\textwidth]{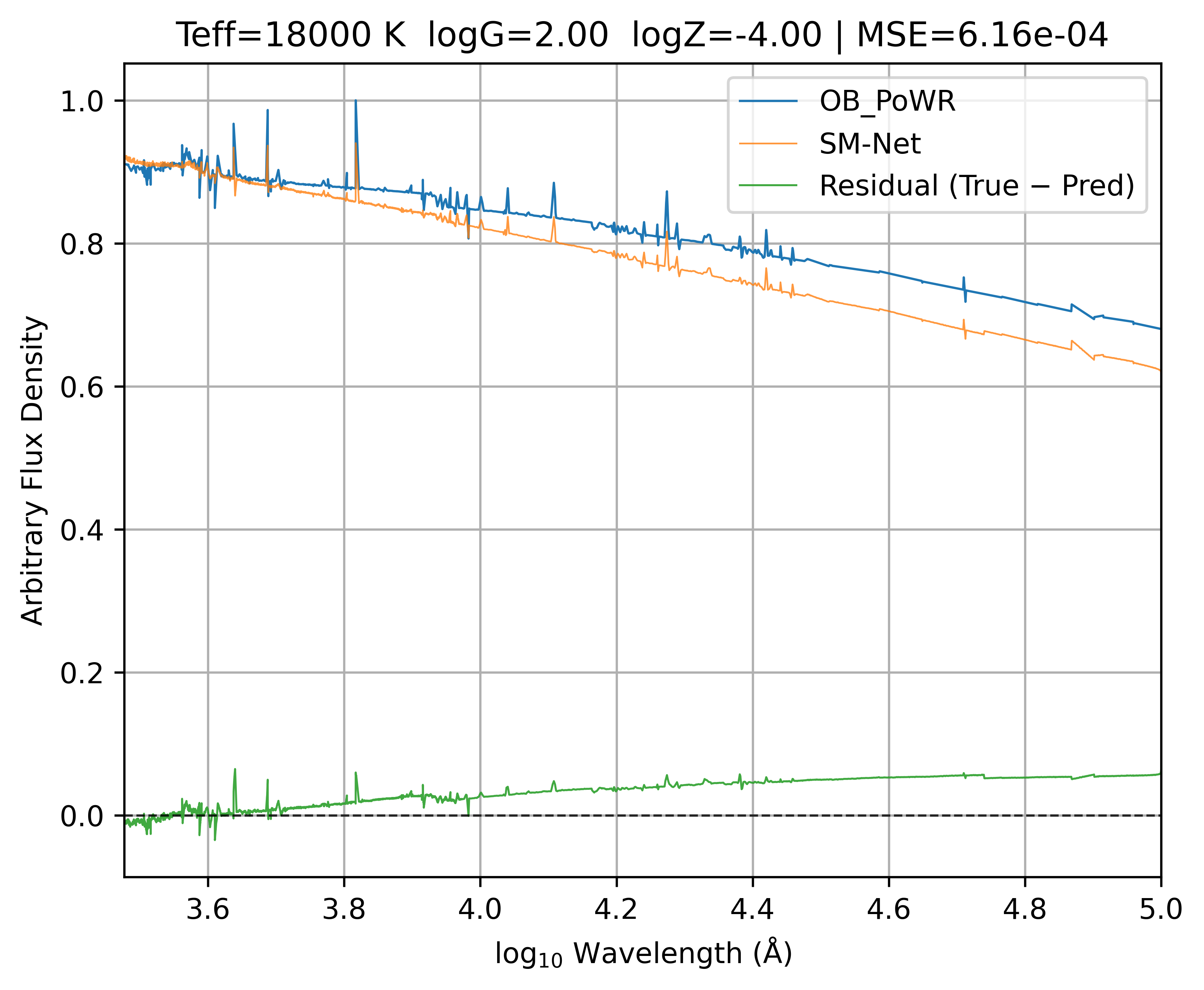}
    \caption{
    Example of a low-quality reconstruction for
    $T_{\mathrm{eff}} = 18,000~\mathrm{K}$, $\log g = 2.0$, and $\log Z = -4.0$,
    compared against the corresponding OB-PoWR reference spectrum.}
    \label{fig:bad2}
\end{figure}

\begin{figure}[!t]
    \centering
    \includegraphics[width=1\textwidth]{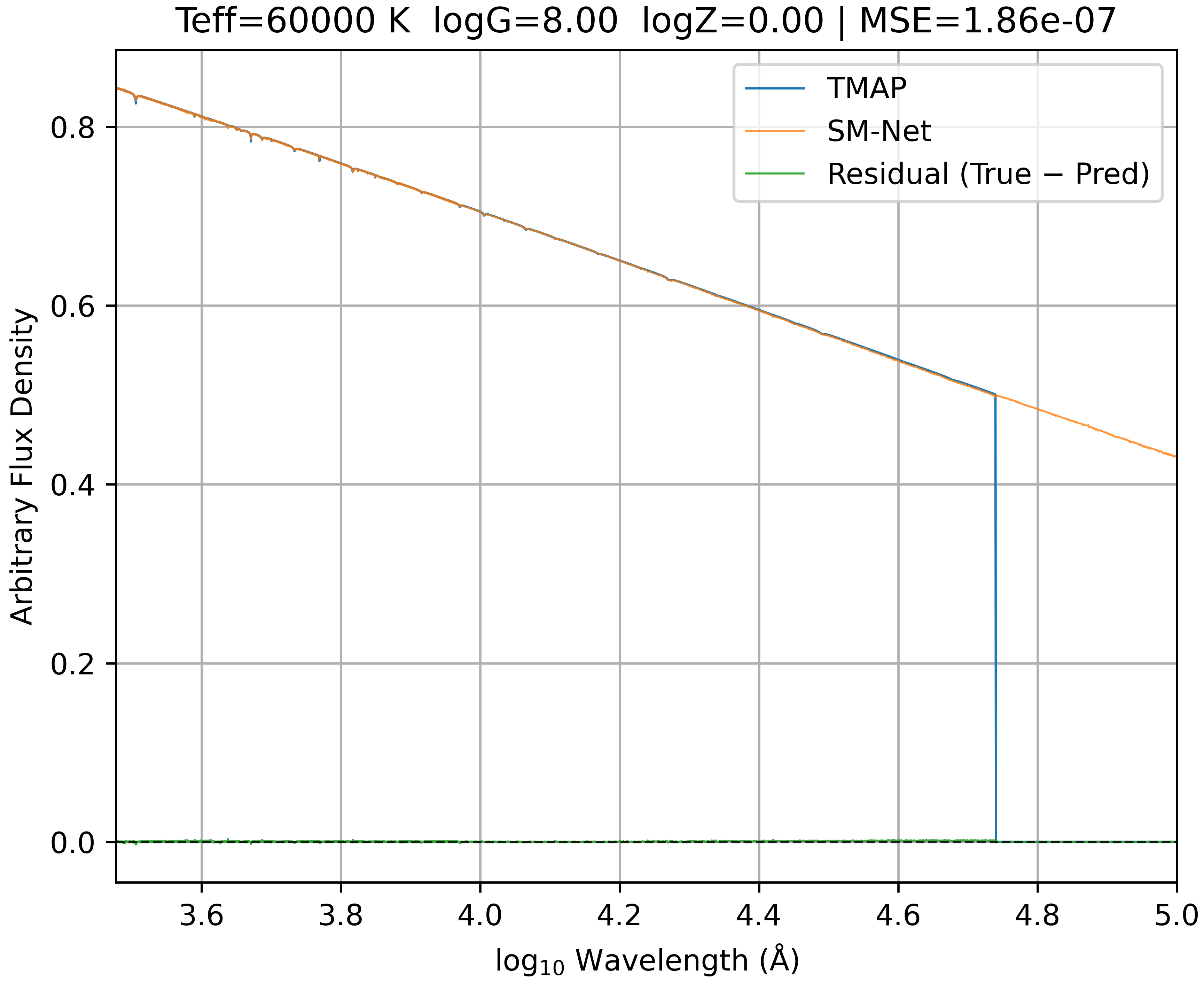}
    \caption{
    Example of SM-Net reconstructing missing high-wavelength flux for
    $T_{\mathrm{eff}} = 60{,}000~\mathrm{K}$ and $\log g = 8.0$,
    compared with the corresponding TMAP reference spectrum.}
    \label{fig:recon}
\end{figure}

For $T_{\mathrm{eff}}$, the per-library breakdown confirms that the largest reconstruction errors at low temperature are concentrated in the \textsc{PHOENIX}--Husser and C3K libraries, while a small number of outliers near $20{,}000\mathrm{K}$ arise mainly from OB-PoWR. At higher temperatures, where OB-PoWR and \textsc{TMAP} dominate, the MSE remains consistently low and outliers are largely absent, indicating stable reconstructions in these regimes. For $\log g$, the highest mean errors occur around $\log g \simeq 2.0$--2.2, primarily within the OB-PoWR regime, while C3K contributes most of the outliers. For $\log Z$, the mean MSE stays uniformly low across all four libraries, with outliers dispersed rather than concentrated in any particular metallicity bin. \textsc{TMAP} is a notable exception, showing virtually no outliers, consistent with its behaviour in the $T_{\mathrm{eff}}$ and $\log g$ distributions.

\subsection{Reconstructed Spectra}

In this section, we present and discuss example spectra reconstructed by SM-Net for our test set. None of the exact parameter combinations for the examples below were available to the model during the training and validation. Also note that the vertical axes for flux and residuals are shown on a logarithmic scale to improve visibility of structures across the wide dynamic range.

Figure \ref{fig:Sun} shows the spectrum in our combined dataset that is closest to the Sun, with parameters $T_{\mathrm{eff}} = 5,800~\mathrm{K}$, $\log g = 4.5$, and $\log Z = 0$. The spectrum predicted by SM-Net closely matches the corresponding PHOENIX-Husser template at the same parameter values. Most of the residual structure appears at shorter wavelengths, but even then the peak deviations remain below $\approx 0.02$ in flux, which is less than 2\% of the dataset’s maximum normalised flux of 1.

Figure \ref{fig:good1} shows the SM-Net reconstruction for a star with
$T_{\mathrm{eff}} = 6,900~\mathrm{K}$, $\log g = 5.0$, and $\log Z = -4.0$.
The model accurately reproduces the fine spectral structure, achieving a mean-squared error of
$5.22 \times 10^{-7}$ relative to the PHOENIX–Husser reference spectrum.
This example illustrates the model’s ability to deliver extremely high-fidelity reconstructions across the combined dataset.

Figure \ref{fig:good2} presents another high-quality reconstruction, this time for a hotter star with $T_{\mathrm{eff}} = 13{,}000~\mathrm{K}$, $\log g = 4.5$, and $\log Z = -1.5$.
SM-Net again reproduces the detailed spectral structure with a mean-squared error of $6.25 \times 10^{-7}$ relative to the corresponding C3K reference spectrum.
This example further demonstrates that SM-Net does not overfit to any single library, but instead adapts smoothly to the templates present in the local region of parameter space.

Figure \ref{fig:bad1} shows an example of a poor-quality reconstruction for $T_{\mathrm{eff}} = 2,200~\mathrm{K}$, $\log g = 3.0$, and $\log Z = -2.1$,
compared against the corresponding C3K reference spectrum. The model exhibits substantial residuals across the full wavelength range, with the largest deviations occurring at shorter wavelengths.
This case highlights the greater difficulty of accurately reconstructing spectra in the very low–temperature regime.

Figure~\ref{fig:bad2} presents another example of a poor-quality reconstruction for
$T_{\mathrm{eff}} = 18{,}000$~$\mathrm{K}$, $\log g = 2.0$, and $\log Z = -4.0$,
compared with the corresponding OB-PoWR reference spectrum.
While SM-Net reproduces the overall shape reasonably well, it struggles to capture the correct slope as a function of wavelength, and the error exceeds 5\% at longer wavelengths.
This example illustrates the challenges associated with modelling spectra where sparse grid coverage makes interpolation particularly difficult.

Figure \ref{fig:recon}, corresponding to $T_{\mathrm{eff}} = 60{,}000~\mathrm{K}$ and $\log g = 8.0$, showcases SM-Net’s ability to reconstruct spectra even when parts of the flux are missing in the original library template. Compared with the TMAP reference spectrum, the model successfully infills the absent high-wavelength flux while maintaining consistency with the surrounding spectral structure.
Such reconstruction is beyond the capability of standard interpolation methods, which cannot recover structure in regions not sampled by the underlying libraries.

Across the full set of examples, SM-Net demonstrates consistently strong performance.
The high-quality cases, ranging from Sun-like stars to hot and moderately metal-poor systems, highlight the model’s ability to capture both the global continuum shape and fine absorption features with MSE values at the $10^{-7}$ level.
The contrasting low-quality examples highlight the inherent difficulty of reconstructing spectra in sparsely sampled or physically extreme regions of parameter space, particularly at very low temperatures where molecular features become complex, and library coverage is limited.
Finally, the reconstruction of spectra with missing flux in the library templates illustrates a key advantage of SM-Net over classical interpolation techniques: its ability to infer plausible spectral structure in the absence of complete training data.

\begin{figure*}[!t]
    \centering
    \includegraphics[width=1\textwidth]{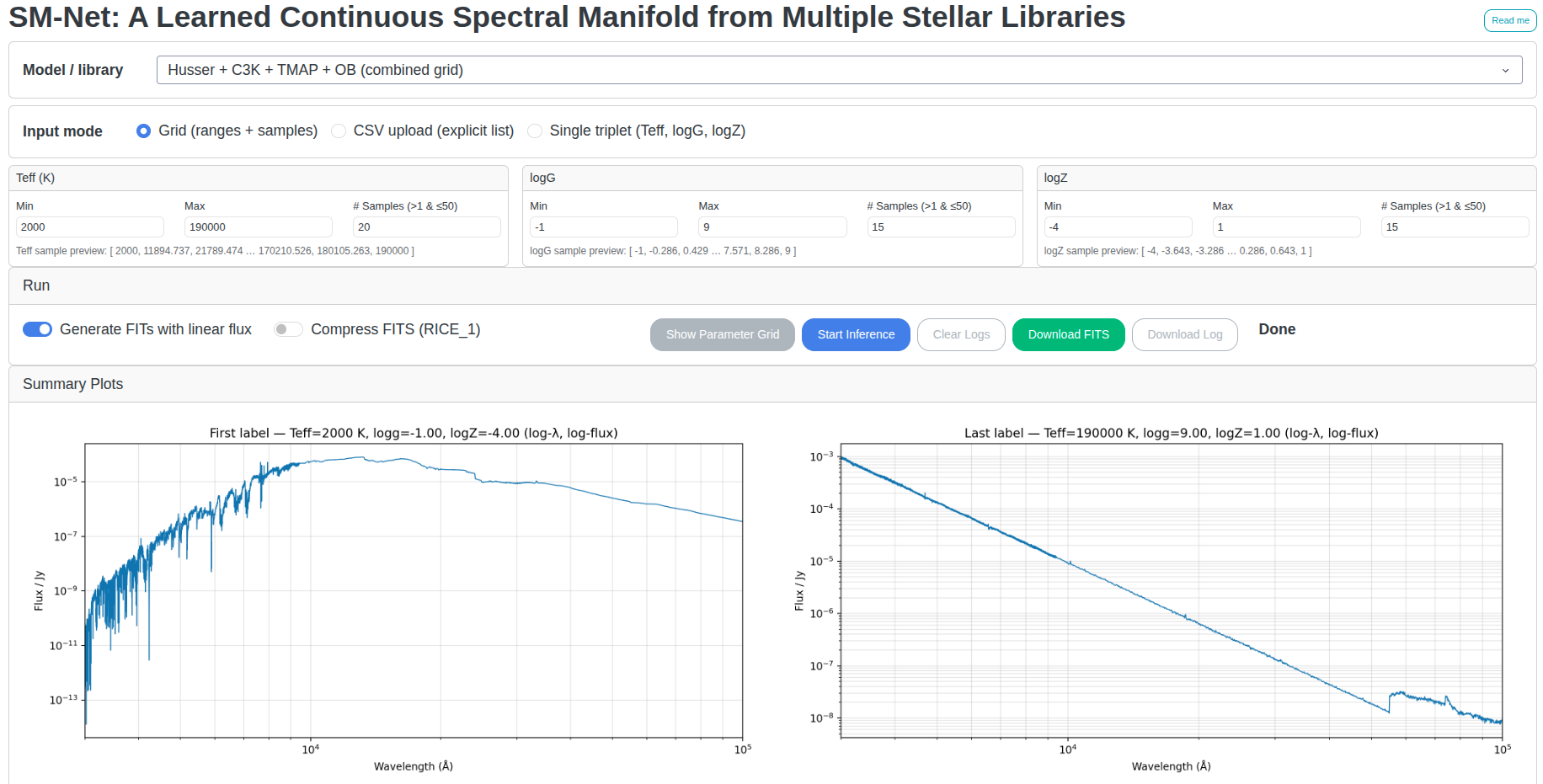}
    \caption{
    The layout of the interactive interface.}
    \label{fig:interface}
\end{figure*}

SM-Net can also be evaluated at parameter combinations outside the convex hull of the training data, producing smooth extrapolative predictions based on trends learned from neighbouring regions of parameter space. However, these predictions are not directly validated by reference spectra and should be treated as exploratory rather than physically established. Consequently, extrapolated spectra should be interpreted with caution and regarded as hypotheses generated by the learned manifold rather than as physically validated models. Systematic assessment of the physical reliability of extrapolated predictions is left to future work.


\section{Interactive Interface}

To promote accessibility and community use, we release SM-Net as an open, self-hosted toolkit via GitHub, including the trained model weights, wavelength/meta files, and an interactive Dash/Plotly dashboard for spectral generation and exploration \citep{smnet_dashboard}. This allows users to run the interface locally or on institutional resources after a simple installation step.

\begin{figure}[!b]
    \centering
    \includegraphics[width=1\textwidth]{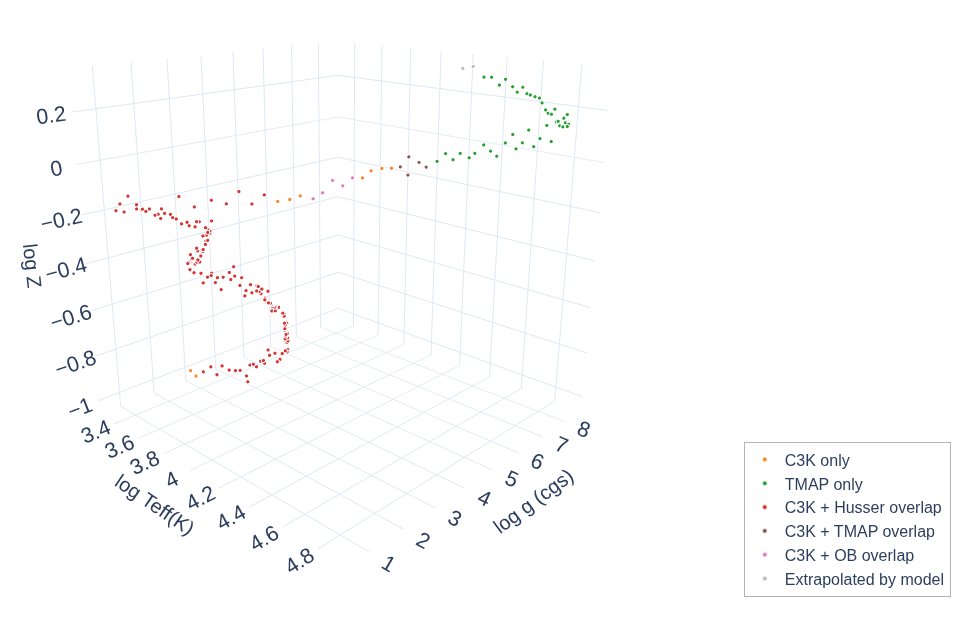}
    \caption{
    Example of a visualisation of a 3D grid via SM-Net interface, based on a CSV uploaded by the user.}
    \label{fig:csv_grid}
\end{figure}

The dashboard provides an interactive workflow for specifying parameters, generating spectra, and inspecting outputs. Users can define the physical parameters $T_{\mathrm{eff}}$, $\log g$, and $\log Z$ through configurable grids, direct numerical inputs, or uploaded label CSVs (Figure~\ref{fig:interface}). Each query produces downloadable spectral outputs together with immediate visual previews. By packaging the full inference stack, the release abstracts away details such as normalisation, wavelength handling, and file formatting, while preserving transparency and reproducibility. The release also includes the wavelength grids, scaling parameters, and associated metadata used for inference, ensuring consistency with the training and evaluation setup.

Beyond spectral generation, the dashboard includes interactive tools for visualising the coverage of the underlying stellar libraries and the combined grid learned by SM-Net. A 3D parameter-space viewer (Figure~\ref{fig:csv_grid}) displays the sampled $(T_{\mathrm{eff}}, \log g, \log Z)$ points, colour-coded by contributing library and overlap region, allowing users to identify interpolation and extrapolation regimes at a glance. Generated FITS cubes can then be used directly in downstream analyses, including population synthesis and spectral fitting workflows.

The dashboard can also be deployed on shared institutional compute resources and accessed through a web browser by multiple users within the same network. This provides a centrally managed and reproducible environment for collaborative use, without requiring separate installation on individual machines. Collectively, this GitHub-based release provides a practical, research-grade pathway for community experimentation and rapid hypothesis testing without requiring a centrally hosted web service.

\section{Discussion}


SM-Net provides an accurate and computationally efficient framework for generating stellar spectra. Across the combined PHOENIX--Husser, C3K--Conroy, OB--PoWR, and TMAP--Werner libraries, it achieves mean RMS flux errors of $\sim 0.38$--$0.51\%$ on the training, validation, and test sets, and performs robust interpolation at parameter combinations not encountered during training. Inference is performed in a single forward pass. On a single NVIDIA RTX 4090 GPU, the model achieves a throughput of approximately 14,000 spectra per second under high-throughput settings. This makes SM-Net well suited to applications requiring dense sampling of stellar parameter space, rapid spectral generation, or interactive exploration of the learned manifold.

A key feature emerging from the learned manifold is the model’s ability to infer structure in masked or zero-valued flux regions.
Because SM-Net treats masked values as unknown rather than physically meaningful zeros, it reconstructs missing spectral information using correlations learned across neighbouring $(T_{\mathrm{eff}}, \log g, \log Z)$ points.
The examples presented in Section~6 demonstrate that the network draws on the smoothness of spectral morphology in local regions of parameter space, allowing it to interpolate through gaps created by incomplete line lists or numerical artefacts in the original libraries.
This behaviour is physically intuitive: neighbouring points in stellar parameter space share similar continuum shapes, ionisation balances, and line strengths, and the network leverages these relationships to fill missing structure in a manner that is often inaccessible to standard grid interpolation.

An important distinction should be made between SM-Net’s ability to interpolate within the joint parameter space spanned by the training libraries and its ability to extrapolate beyond that space. Interpolation performance is directly validated against reference spectra and constitutes the primary result of this work. Extrapolation, while supported by the continuity of the learned manifold, is not constrained by ground-truth data and should therefore be treated as exploratory. This behaviour is a natural consequence of learning a smooth functional mapping and does not imply physical correctness outside the domain sampled by the input libraries.

Despite its strengths, the present version of SM-Net also inherits several limitations from the training data and model design.
First, the emulator predicts static spectra without velocity broadening or instrumental convolution.
These effects must be applied downstream, analogous to the workflow used in stellar population synthesis codes.
Second, the model learns the assumptions embedded in each training library, including differences in opacities and line lists across grid overlaps.
Residual structure at the boundaries between PHOENIX–Husser and C3K, for example, reflects genuine inconsistencies between the physical models rather than network failure.
Third, the wavelength range is constrained to the adopted 3,000--100,000~\AA\ grid; behaviour outside this interval requires the inclusion of training spectra covering additional wavelength ranges with physical consistencies. 

A critical limitation arises in the high-temperature regime ($T_{\mathrm{eff}} > 15,000$\,K), which is dominated by the \textsc{OB-PoWR} and \textsc{TMAP} libraries. Because these training grids do not strictly parametrise metallicity, SM-Net produces spectra that are mostly invariant to $\log Z$ in these regions. 
Users should treat this as a regime where metallicity is unconstrained by the underlying physical models. The model effectively marginalises over metallicity for hot stars; therefore, SM-Net should not be used to infer metallicities or study metallicity-dependent features (such as wind lines or line blanketing) for O-type, B-type, or White Dwarf stars. 
In these regimes, the $\log Z$ input acts merely as a dummy variable required for the network's architectural consistency and does not represent a physical prediction of metallicity effects.

These considerations highlight opportunities for hybrid approaches combining machine-learning emulation with physical radiative-transfer modules.
For example, an ML component could provide rapid coarse predictions or act as a differentiable prior, while a physics-based solver refines selected wavelength regions or enforces specific constraints (e.g.\ ionisation equilibrium).
Such hybrid models would retain physical interpretability while leveraging the efficiency and smooth interpolation capabilities of data-driven emulators.
SM-Net therefore represents both a proof of concept and a foundation for next-generation spectral synthesis frameworks that balance physical rigour with computational scalability.

\section{Conclusion and Future Work}

SM-Net provides a data-driven framework for representing the composite stellar-library dataset as a smooth and easily navigable mapping in parameter space. Rather than replacing existing libraries or physical modelling pipelines, it acts as an integration layer over the constructed training manifold, mitigating the practical effects of irregular grid sampling and enabling efficient spectral generation across the combined parameter domain. This supports analysis modes that are cumbersome with discrete template grids, including dense parameter sweeps, interactive exploration of stellar parameter space, and rapid generation of stellar spectra for downstream modelling.

A key practical strength of SM-Net is its accessibility and reproducibility.
All stages of the workflow, including dataset construction, preprocessing, training, and evaluation, are fully reproducible.
The browser-based Dash demonstrator further lowers the barrier to use, allowing researchers to interrogate the learned manifold, examine cross-library behaviour, and generate spectra for arbitrary parameter combinations without requiring specialised computational resources or bespoke interpolation code.

Looking ahead, the most natural extensions of this work lie in stellar population synthesis rather than in individual stellar atmosphere libraries.
A closely related manifold-learning approach could be applied directly in simple stellar population (SSP) space by generating large ensembles of population models with controlled variations in ingredients such as the initial mass function, isochrones, or underlying stellar libraries.
Learning a continuous representation in this higher-level parameter space would enable efficient exploration of population-scale dependencies while maintaining a clear separation between stellar-atmosphere physics and population-synthesis assumptions.

It is also important to place SM-Net in the context of existing interpolation-based approaches.
Classical schemes, such as the MSG interpolator \citep{townsend2023msg} and the methods described in the ProGeny-I framework, provide physically motivated interpolation within discrete grids and form a valuable baseline for stellar spectral modelling.
SM-Net offers a complementary alternative that is optimised for scalability, smoothness across heterogeneous libraries, and rapid inference, particularly in regimes where grid irregularities or library boundaries limit the effectiveness of traditional interpolators.

Beyond stellar atmospheres, the same manifold-learning framework could be extended to other forward-modelling tasks, including nebular emission models, binary-evolution tracks, or machine-learning-accelerated components of radiative-transfer workflows.
The model could also be coupled to simple physical transformations, such as velocity broadening or instrumental response functions, to enable more direct integration with observational analysis pipelines.

In summary, SM-Net provides a flexible and efficient foundation for next-generation spectral modelling.
By unifying heterogeneous stellar libraries into a coherent continuous manifold, it bridges detailed stellar atmosphere calculations and scalable analysis workflows, supporting both traditional population-synthesis applications and new, interactive approaches to exploring stellar parameter space.
Its extensibility positions it well for integration into future hybrid physical--machine-learning pipelines and for meeting the demands of large-scale spectroscopic surveys in the coming decade.

\section*{Acknowledgements}

We would like to extend our sincere thanks to the Progeny team for generously sharing the libraries used for training, testing and validating the model. Luca Cortese acknowledges support from the Australian Research Council Discovery Project funding scheme (DP210100337). Aaron Robotham acknowledges support from the Australian Research Council Future Fellowship (FT200100375).

\clearpage

\end{document}